\documentclass[aps,pre,onecolumn,floatfix,longbibliography]{revtex4}

\usepackage[T1]{fontenc} 
\usepackage{color}
\usepackage[dvipsnames]{xcolor}
\usepackage[english]{babel} 
\usepackage{amsmath,amsfonts,amsthm} 
\usepackage{graphicx}
\usepackage{slashed}
\usepackage{dsfont}
\usepackage{extramarks} 
\usepackage{fancyhdr}
\usepackage{caption}
\usepackage{hyperref}
\allowdisplaybreaks

\usepackage{epsf}
\usepackage{epstopdf}
\usepackage{mathrsfs}

\usepackage{soul}
\usepackage{ulem}

\numberwithin{equation}{section} 
\numberwithin{figure}{section}
\numberwithin{table}{section} 

\newcommand{\be}{\begin{equation}}
\newcommand{\ee}{\end{equation}}
\newcommand{\bea}{\begin{eqnarray}}
\newcommand{\eea}{\end{eqnarray}}

\begin{document}
\setlength{\unitlength}{1mm}

\title{Nonequilibrium statistical mechanics of crystals}

\author{Jo\"el Mabillard}
\email{joel.mabillard@ulb.ac.be; ORCID: 0000-0001-6810-3709.}
\author{Pierre Gaspard}
\email{gaspard@ulb.ac.be; ORCID: 0000-0003-3804-2110.}
\affiliation{Center for Nonlinear Phenomena and Complex Systems, Universit{\'e} Libre de Bruxelles (U.L.B.), Code Postal 231, Campus Plaine, B-1050 Brussels, Belgium}

\vskip 0.5 cm

\begin{abstract}
The local equilibrium approach previously developed by the Authors [J. Mabillard and P.~Gaspard, {\it J. Stat. Mech.} (2020) 103203] for matter with broken symmetries is applied to crystalline solids.  The macroscopic hydrodynamics of crystals and their local thermodynamic and transport properties are deduced from the microscopic Hamiltonian dynamics.  In particular, the Green-Kubo formulas are obtained for all the transport coefficients.  The eight hydrodynamic modes and their dispersion relation are studied for general and cubic crystals.  In the same twenty crystallographic classes as those compatible with piezoelectricity, cross effects coupling transport between linear momentum and heat or crystalline order are shown to split the degeneracy of damping rates for modes propagating in opposite generic directions.
\end{abstract}


\maketitle


\section{Introduction}
\label{Introduction}

Crystals manifest long-range order by the spatial periodicity of their atomic structure, which can be classified into 14 Bravais lattices, 32 crystallographic point groups, and 230 space groups \cite{AM76,K76,IL09}. In contrast, fluids have uniform and isotropic properties and are thus symmetric under continuous spatial translations and rotations.  According to Goldstone's theorem, the breaking of the three-dimensional continuous group of spatial translations in crystals implies the existence of three slow modes, called Nambu-Goldstone modes, in addition to the five slow modes arising from the fundamental conservation laws of mass, energy, and linear momentum \cite{N60,G61,S66,A84}.  In bulk matter, each mode is characterized by a dispersion relation between its frequency (or rate) and its wave number (or wave length).  The slow modes, also called hydrodynamic modes, are characterized by the vanishing of their dispersion relation with the wave number.  The mode is propagative (respectively, diffusive) if the dispersion relation vanishes linearly (respectively, quadratically) with the wave number.  As a consequence of continuous symmetry breaking, there exist eight hydrodynamic modes in crystals: two longitudinal sound modes, four transverse sound modes, one heat mode, and an additional mode, which has been identified as the mode of vacancy diffusion \cite{MPP72}.  With these considerations, the macroscopic hydrodynamics of crystals is well established since the seventies \cite{MPP72,FC76}.  Earlier treatments of hydrodynamics in solids had identified seven modes only, because the atoms were supposed to move without leaving their lattice cell.  However, thermal fluctuations may induce the motion of atoms out of their lattice cell, yielding point defects called vacancies and interstitials in the crystal \cite{AM76,K76,IL09}, which is the mechanism of generating the vacancy diffusion mode.

If the diffusive parts of the dispersion relations are neglected, the modes are either propagative or static, so that there is no energy dissipation and entropy is conserved.  Instead, the diffusivity of the modes is associated with the transport coefficients, which are responsible for damping and irreversible entropy production.  An issue of paramount importance is to deduce these macroscopic properties from the underlying microscopic motion of atoms, which is the programme of nonequilibrium statistical mechanics.

Since the fifties, this programme is carried out for normal fluids, starting from the microscopic expressions for the densities of the five fundamentally conserved quantities and their time evolution according to Hamiltonian microdynamics \cite{M58,KM63} and using the local equilibrium approach \cite{McL60,McL61,McL63,DMcL65,Z66,R66,R67,P68,DK72,OL79,BZD81,KO88,BY80,AP81,Sp91,S14} often combined with the projection-operator method \cite{Z61,M65}.  In this way, Green-Kubo formulas can be deduced for the transport coefficients in normal fluids \cite{G52,G54,K57,H60}.  In order to extend this programme to matter phases with broken continuous symmetries, the associated order parameter must be identified in the microscopic description of these phases \cite{F74,F75,CL95}.

Since the nineties, the microscopic expression is known for the crystalline order parameter, which is the displacement vector field, providing the statistical-mechanical formulation of hydrodynamics in crystals \cite{SE93,S97}.  Recently, this formulation has led to a systematic study of elastic properties in nonideal crystals, i.e., crystals with point defects \cite{WF10,HWSF15}.  In addition to elasticity, the transport properties of nonideal crystals can also be investigated. Using the local equilibrium approach, the Authors have recently developed a unified statistical-mechanical theory for hydrodynamics in phases with broken symmetries, in particular, deducing all the Green-Kubo formulas for the transport coefficients from the microdynamics \cite{MG20}.  In certain non-centrosymmetric anisotropic phases, this systematic approach combined with Curie's principle~\cite{C1894} and Onsager-Casimir reciprocal relations~\cite{O31a,O31b,C45} predicts the existence of cross effects coupling the transport of linear momentum to those of heat and the order emerging from symmetry breaking.

In the present paper, our purpose is to apply the systematic approach of reference~\cite{MG20} to crystals in order to obtain their thermodynamic and transport properties.  The consequences of the crystallographic symmetries on the transport properties are investigated.  Transport coefficients are identified forming tensors of rank two, three, and four.  Furthermore, the dispersion relations of the eight hydrodynamic modes are obtained for crystallographic classes where rank-three tensors are or are not vanishing.

The plan of the paper is the following.  Section~\ref{Sec:Micro} is devoted to the formulation of the microscopic Hamiltonian dynamics and the construction of the crystalline order parameter.
Section~\ref{Sec:NESM} presents the main results of reference~\cite{MG20} about the local equilibrium approach for the nonequilibrium statistical mechanics of crystals and the deduction of their hydrodynamic equations.  The eight hydrodynamic modes and their dispersion relation are obtained in section~\ref{Sec:Macro} for general crystals and in section~\ref{Sec:Cubic} for cubic crystals.  Section~\ref{Sec:Conclusion} is concluding the paper.

\section{Microscopic description}
\label{Sec:Micro}

In this section, the Hamiltonian microdynamics is formulated for crystals, the densities obeying local conservation laws are introduced at the microscopic level of description, and the crystalline order parameter is constructed on the basis of symmetry breaking from continuous to discrete spatial translations.

\subsection{The Hamilton equations of motion}

At the microscopic scale, the crystal is composed of $N$ atoms of mass $m$.  The atoms have the positions and momenta $\Gamma=({\bf r}_i,{\bf p}_i)_{i=1}^N\in{\mathbb R}^{6N}$.  They are mutually interacting by the energy potential $V(r_{ij})$ with $r_{ij}=\Vert{\bf r}_i-{\bf r}_j\Vert$, so that the total energy of the crystal is given by the following Hamiltonian function,
\be\label{Hamilt_fn}
H = \sum_i \frac{{\bf p}_i^2}{2m} + \frac{1}{2}\sum_{i\neq j} V(r_{ij}) \, .
\ee
Consequently, the motion of the atoms is governed by Hamilton's equations
\be\label{Hamilt_eqs}
\frac{{\rm d}r_i^a}{{\rm d}t} = \frac{\partial H}{\partial p_i^a} = \frac{p_i^a}{m} \, , \qquad\qquad
\frac{{\rm d}p_i^a}{{\rm d}t} = -\frac{\partial H}{\partial r_i^a} = \sum_{j(\ne i)} F_{ij}^a \, ,
\ee
where
\be
F^a_{ij}\equiv - \frac{\partial V(r_{ij})}{\partial r_i^a}
\ee
is the $a^{\rm th}$ component of the interaction force exerted on the $i^{\rm th}$ atom by the $j^{\rm th}$ atom (for $a=x,y,z$ and $i,j=1,2,\dots,N$).  Since the Hamiltonian function is invariant under spatiotemporal translations and rotations, this dynamics is known to conserve the total energy $E=H$, the total momentum ${\bf P}=\sum_i{\bf p}_i$, and the total angular momentum ${\bf L}=\sum_i{\bf r}_i\times{\bf p}_i$, in addition to the total mass $M=mN$.  We note that, for computational simulation with molecular dynamics, the $N$-body system can be considered on a torus with periodic boundary conditions \cite{FS02,AT17}.

\subsection{Locally conserved quantities}

In order to describe the local properties in the crystal, the mass density is defined as $\hat\rho=m\hat n$ in terms of the particle density
\be
\hat n({\bf r};\Gamma) \equiv \sum_i \delta({\bf r}-{\bf r}_i) \, ,
\ee
the energy density as
\be
\hat e({\bf r};\Gamma) \equiv \sum_i E_i \, \delta({\bf r}-{\bf r}_i) \qquad\mbox{with}\qquad E_i\equiv \frac{{\bf p}_i^2}{2m} + \frac{1}{2}\sum_{j (\ne i)} V(r_{ij}) \, ,
\ee
and the momentum density as
\be
\hat g^a({\bf r};\Gamma) \equiv \sum_i p_i^a \, \delta({\bf r}-{\bf r}_i) \, .
\ee
Integrating these densities over the volume of the system gives the total mass, energy, and momentum, which are conserved by the microdynamics.

Using Hamilton's equations~(\ref{Hamilt_eqs}), the aforedefined densities can be shown to obey the following local conservation equations,
\bea
&&\partial_t\hat{\rho}  +  \nabla^a \hat{J}_\rho^{a} = 0\;, \label{eq-rho}\\
&&\partial_t\hat{e}  +  \nabla^a \hat{J}_e^{a} = 0\;, \label{eq-e}\\
&&\partial_t\hat{{g}}^b  +  \nabla^a \hat{J}_{g^b}^{a} = 0\; , \label{eq-g}
\eea
expressed with the corresponding current densities or fluxes,
\bea
&&\hat{{J}}_\rho^{a}(\mathbf{r};\Gamma) \equiv \hat{g}^a(\mathbf{r};\Gamma) \; , \\
&&\hat{{J}}_e^{a}(\mathbf{r};\Gamma) \equiv \sum_i \frac{p_i^a}{m}\, E_i \, \delta(\mathbf{r} - \mathbf{r}_i)  + \frac{1}{2}\sum_{i<j}({r}_i^a - r_j^a)\, \frac{{p}^b_i + {p}^b_j}{m}\,  {F}^b_{ij}\, D(\mathbf{r};\mathbf{r}_i,\mathbf{r}_j) \; , \\
&&\hat{{J}}_{g^b}^{a}(\mathbf{r};\Gamma) \equiv \sum_{i}\frac{p_i^a}{m}\, p_i^b \, \delta(\mathbf{r} - \mathbf{r}_i) + \sum_{i<j} ({r}_i^a - r_j^a) \, {F}^b_{ij} \, D(\mathbf{r};\mathbf{r}_i,\mathbf{r}_j) \; ,
\eea
where
\be
D(\mathbf{r};\mathbf{r}_i,\mathbf{r}_j) \equiv \int_{0}^1 {\rm d}\xi\ \delta\left[\mathbf{r} - \mathbf{r}_i + (\mathbf{r}_i - \mathbf{r}_j)\xi\right]
\ee
is a lineal distribution joining the positions ${\bf r}_i$ and ${\bf r}_j$ \cite{R67,P68,Sp91,S14}.

\subsection{Continuous symmetry breaking into a crystallographic group}

In crystals, the continuous symmetry of the microdynamics under spatial translations,
\be\label{space_transl}
T^{\bf R} A[({\bf r}_i,{\bf p}_i)_{i=1}^N] = A[({\bf r}_i-{\bf R},{\bf p}_i)_{i=1}^N]  \qquad\mbox{with}\qquad {\bf R}\in{\mathbb R}^3
\ee
acting on some observable quantity $A$, is broken into one of the 230 discrete crystallographic space groups.  This continuous symmetry breaking generates long-range order in the crystal, whereupon the mass density becomes spatially periodic and anisotropic in the equilibrium crystalline phase, although the mass density is uniform and isotropic in the fluid phase.  If the translation vector $\bf R$ is infinitesimal, the transformation~(\ref{space_transl}) can be expressed as
\be\label{inf_space_transl}
T^{\bf R} A = A + {\bf R}\cdot \{ {\bf P}, A\} + O({\bf R}^2) 
\ee
in terms of the Poisson bracket $\{\cdot,\cdot\}$ with the total momentum ${\bf P}=\sum_i {\bf p}_i$.

{\it A priori}, the equilibrium properties could be described by the Gibbs grand canonical distribution
\be\label{equil-grd-can-0}
p_{{\rm eq}}(\Gamma) = \frac{1}{\Xi \, \Delta\Gamma} \, {\rm e}^{-\beta (H-\mu M) } \, , 
\ee
where $H$ is the Hamiltonian function~(\ref{Hamilt_fn}), $M=mN$, $\beta=(k_{\rm B}T)^{-1}$ the inverse temperature, $\mu$ the chemical potential, $\Delta\Gamma=h^{3N}$ ($h$ being Planck's constant), and $\Xi$ the partition function such that the normalization condition $\int p_{{\rm eq}}(\Gamma)\, {\rm d}\Gamma =1$ is satisfied with
\be\label{normalization}
\int (\cdot)\, {\rm d}\Gamma = \sum_{N=0}^{\infty} \frac{1}{N!} \int_{{\mathbb R}^{6N}} (\cdot) \prod_{i=1}^N {\rm d}{\bf r}_i\, {\rm d}{\bf p}_i \, .
\ee
The issue is that the probability distribution~(\ref{equil-grd-can-0}) is invariant under the spatial translations~(\ref{inf_space_transl}) because $\{ {\bf P},H\}=0$ and $\{ {\bf P},M\}=0$, although the continuous translational symmetries are broken in the crystalline phase.

In order to induce symmetry breaking, an external energy potential $V^{\rm (ext)}({\bf r})$ may be introduced, so that the Hamiltonian function is modified into
\be
H_{\epsilon} = H+\epsilon \, H^{\rm (ext)} \qquad\mbox{with}\qquad H^{\rm (ext)}=\int V^{\rm (ext)}({\bf r})\, \hat n({\bf r};\Gamma)\, {\rm d}{\bf r} \, .
\ee
Unless the external potential is uniform, the associated grand canonical probability distribution
\be\label{equil-grd-can}
p_{{\rm eq},\epsilon}(\Gamma,N) = \frac{1}{\Xi_{\epsilon} \, \Delta\Gamma} \, {\rm e}^{-\beta (H_{\epsilon}-\mu M) }
\ee
is no longer invariant under the spatial translations~(\ref{inf_space_transl}).  Therefore, taking the mean value of equation~(\ref{inf_space_transl}) for the particle density $A=\hat n({\bf r};\Gamma)$ over the distribution~(\ref{equil-grd-can}) gives
\be
\langle \{ {\bf P}, \hat n({\bf r};\Gamma)\} \rangle_{{\rm eq},\epsilon} = \partial_{\bf r} \langle \hat n({\bf r};\Gamma) \rangle_{{\rm eq},\epsilon} \ne 0 \qquad\mbox{if} \qquad \epsilon\ne 0 \, ,
\ee
because of the explicit symmetry breaking by the external potential.  Under pressure and temperature conditions where the crystalline phase is thermodynamically stable, the limit $\epsilon\to 0$ could be taken in order to remove the external potential and the periodic equilibrium density of the crystal would be obtained as
\be
n_{\rm eq}({\bf r}) \equiv \lim_{\epsilon\to 0} \langle \hat n({\bf r};\Gamma) \rangle_{{\rm eq},\epsilon} \, .
\ee
In this case, the crystalline phase emerges by spontaneous symmetry breaking and its long-range order is characterized by the periodic equilibrium density $n_{\rm eq}({\bf r})$.  Accordingly, the spatial structure formed by the atoms becomes periodic in space. Nevertheless, the center of mass ${\bf R}_{\rm cm}\equiv (1/N)\sum_{i=1}^N{\bf r}_i$ of the whole crystal can undergo spatial translations.  In this regard, we note that  the Hamiltonian function~(\ref{Hamilt_fn}) can be expressed as $H={\bf P}^2/(2M)+H_{\rm rel}$ in terms of the total kinetic energy of the crystal center of mass and the Hamiltonian function~$H_{\rm rel}$ ruling the motion of its atoms relative to the center of mass.  In the symmetric grand canonical ensemble~(\ref{equil-grd-can-0}), the center of mass is uniformly distributed in space with a Maxwellian distribution of its velocity, so that its  trajectories are free flights, ${\bf R}_{\rm cm}(t)={\bf R}_{\rm cm}(0)+t {\bf P}/M$, and the symmetry can only be broken in the frame moving with the center of mass.  In the presence of such spontaneous breaking of continuous symmetry,  there should exist some external force fields exerting arbitrarily small changes in the total energy of the crystal.  In general, an external force field can be described by the external energy potential $V^{\rm (ext)}({\bf r})$, so that the total energy is changed by the mean external energy
\be
\langle H^{\rm (ext)}\rangle_{{\rm eq},0} = \int V^{\rm (ext)}({\bf r}) \, n_{\rm eq}({\bf r}) \, {\rm d}{\bf r}
\ee
in the limit $\epsilon\to 0$. Remarkably, this mean external energy is strictly equal to zero  if the external potential is taken as
\be\label{uniform_V_ext}
V^{\rm (ext)}({\bf r}) = {\bf C}\cdot \pmb{\nabla} n_{\rm eq}({\bf r})
\ee
for some constant vector $\bf C$.  Indeed, for this energy potential, an integration by parts leads to
\be
\langle H^{\rm (ext)}\rangle_{{\rm eq},0} = \int {\bf C}\cdot \pmb{\nabla} n_{\rm eq}({\bf r}) \, n_{\rm eq}({\bf r}) \, {\rm d}{\bf r} = - \int n_{\rm eq}({\bf r}) \, {\bf C}\cdot \pmb{\nabla} n_{\rm eq}({\bf r}) \, {\rm d}{\bf r} = -\langle H^{\rm (ext)}\rangle_{{\rm eq},0} = 0 \, ,
\ee
hence the result. For such external potentials, the perturbation may induce crystal formation, but does not change the total energy with respect to the value of the unperturbed Hamiltonian.  Since its presence costs no energy on average, the external potential~(\ref{uniform_V_ext}) is leading to the construction of the crystalline order parameter, as shown below.

\subsection{Local order parameter in crystals}

At the macroscale, the crystal should be described in terms of macrofields that slowly vary in space on scales larger than the periodic crystalline structure.  This feature can be expressed by requiring that the macrofields have no Fourier mode outside the first Brillouin zone ${\cal B}$.  In general, any function can be decomposed into Fourier modes according to
\bea
f({\bf r}) &=& \int_{{\mathbb R}^3} \frac{{\rm d}{\bf q}}{(2\pi)^3} \, \tilde f({\bf q}) \, {\rm e}^{\imath {\bf q}\cdot{\bf r}} \, , \label{IFT}\\
\tilde f({\bf q}) &=& \int_{{\mathbb R}^3} {\rm d}{\bf r} \, f({\bf r}) \, {\rm e}^{-\imath {\bf q}\cdot{\bf r}} \, . \label{FT}
\eea
For a macrofield $x$, we thus require that
\be\label{macrofield-dfn-q}
\tilde x({\bf q}) = I_{\cal B}({\bf q})\, \tilde x({\bf q}) \, ,
\ee
where $I_{\cal B}({\bf q})$ is the indicator function of the first Brillouin zone of the crystalline lattice.  

We note that, for any arbitrary function $\tilde f({\bf q})$, the transformation $\tilde f({\bf q})\to I_{\cal B}({\bf q})\, \tilde f({\bf q})$ is a projection into the functional space of macrofields.  In the position space, this transformation is expressed as
\be\label{transform}
f({\bf r}) \to \int_{{\mathbb R}^3} \Delta({\bf r}-{\bf r'}) \, f({\bf r'}) \, {\rm d}{\bf r'}
\ee
in terms of the function
\be\label{Delta-dfn}
\Delta({\bf r}-{\bf r'}) \equiv \int_{\cal B} \frac{{\rm d}{\bf k}}{(2\pi)^3} \, {\rm e}^{\imath {\bf k}\cdot({\bf r}-{\bf r'})} \, .
\ee
This function satisfies the property that $\int_{{\mathbb R}^3} \Delta({\bf r}-{\bf r'}) \, {\rm d}{\bf r'} = 1$, as for a Dirac delta distribution.  Moreover, the function~(\ref{Delta-dfn}) is real because the first Brillouin zone is symmetric under the inversion ${\bf q}\to -{\bf q}$.  Indeed, the first Brillouin zone is the Wigner-Seitz primitive cell of the reciprocal lattice, which is a Bravais lattice.  A Wigner-Seitz primitive cell has the full symmetry of its Bravais lattice and the point group of a Bravais lattice always includes the inversion.  Therefore, we have that
\be\label{Delta-real}
\Delta({\bf r}-{\bf r'})^*=\Delta({\bf r'}-{\bf r})=\Delta({\bf r}-{\bf r'}) \, .
\ee
As expected, the transformation~(\ref{transform}) is also a projection because
\be
\Delta({\bf r}-{\bf r'}) = \int_{{\mathbb R}^3} \Delta({\bf r}-{\bf r''}) \, \Delta({\bf r''}-{\bf r'}) \, {\rm d}{\bf r''} \, ,
\ee
as can be shown using the definition~(\ref{Delta-dfn}).  According to these considerations, the property~(\ref{macrofield-dfn-q}) for the macrofield $x$ becomes
\be\label{macrofield-condition}
x({\bf r}) = \int_{{\mathbb R}^3} \Delta({\bf r}-{\bf r'}) \, x({\bf r'}) \, {\rm d}{\bf r'} \, .
\ee

In order to identify the local order parameter associated with the continuous symmetry breaking in crystals, we consider the following external perturbation
\be\label{H-ext-0}
H^{\rm (ext)} = \int V^{\rm (ext)}({\bf r}) \left[\hat n({\bf r};\Gamma)-n_{\rm eq}({\bf r})\right] {\rm d}{\bf r} = \int {\bf C}({\bf r})\cdot \pmb{\nabla} n_{\rm eq}({\bf r}) \, \left[\hat n({\bf r};\Gamma)-n_{\rm eq}({\bf r})\right] {\rm d}{\bf r} \, ,
\ee
obtained by extending the constant vector $\bf C$ in equation~(\ref{uniform_V_ext}) into a macrofield ${\bf C}({\bf r})$ and by substracting the equilibrium value for $\epsilon=0$, so that $\langle H^{\rm (ext)}\rangle_{{\rm eq},0} =0$.  Accordingly, this external perturbation costs arbitrarily small energy in the limit where ${\bf C}({\bf r})$ becomes constant in space.
Since ${\bf C}({\bf r})$  is a macrofield, it satisfies equation~(\ref{macrofield-condition}) with the function~(\ref{Delta-dfn}), whereupon the external perturbation~(\ref{H-ext-0}) can be written in the following form,
\bea
H^{\rm (ext)} &=& \int \int \Delta({\bf r}-{\bf r'}) \, {\bf C}({\bf r'})\cdot \pmb{\nabla} n_{\rm eq}({\bf r}) \left[\hat n({\bf r};\Gamma)-n_{\rm eq}({\bf r})\right] {\rm d}{\bf r} \, {\rm d}{\bf r'} \nonumber\\
&=& \int {\rm d}{\bf r} \, {\bf C}({\bf r})\cdot \int {\rm d}{\bf r'} \, \Delta({\bf r}-{\bf r'}) \, \pmb{\nabla}' n_{\rm eq}({\bf r'}) \left[\hat n({\bf r'};\Gamma)-n_{\rm eq}({\bf r'})\right] ,
\label{H-ext-1}
\eea
using the property~(\ref{Delta-real}).

At the macroscale, such an external perturbation of the crystal would be described as
\be\label{Hext-phi-u}
H^{\rm (ext)} = -\int {\rm d}{\bf r} \, \pmb{\phi}({\bf r}): \hat{\boldsymbol{\mathsf u}}({\bf r};\Gamma)
\ee
in terms of the symmetric tensor $\pmb{\phi}=\pmb{\phi}^{\rm T}$ and the strain tensor
\be\label{strain-dfn}
\hat{\boldsymbol{\mathsf u}} \equiv \frac{1}{2} \left( \pmb{\nabla}\hat{\bf u}+\pmb{\nabla}\hat{\bf u}^{\rm T}\right) ,
\ee
where $\hat{\bf u}({\bf r};\Gamma)$ is the displacement vector field, here supposed to be defined at the microscopic level of description.  The tensor $\pmb{\phi}=(\phi^{ab})$ describes the stress applied to the crystal by the external perturbation because an integration by parts of equation~(\ref{Hext-phi-u}) with the definition~(\ref{strain-dfn}) gives
\be\label{Hext-f-u}
H^{\rm (ext)} =  \int {\rm d}{\bf r}\, (\pmb{\nabla}\cdot\pmb{\phi})\cdot \hat{\bf u} = - \int {\rm d}{\bf r} \, {\bf f}\cdot \hat{\bf u} \, ,
\ee
after introducing the force density field
\be\label{force_density}
{\bf f} \equiv - \pmb{\nabla}\cdot\pmb{\phi} \, .
\ee
Comparing equations~(\ref{H-ext-1}) and~(\ref{Hext-f-u}), we see that the vector field ${\bf C}({\bf r})$ should correspond to the force density field ${\bf f}({\bf r})$ and the rest to the microscopic displacement vector field~$\hat{\bf u}({\bf r};\Gamma)$.  This latter should also satisfy the property that, under a homogeneous dilatation ${\bf r}\to\lambda{\bf r}$ of the lattice by a factor $\lambda$, the mean value of the displacement vector field should be given by $\langle\hat{\bf u}({\bf r};\Gamma)\rangle = (\lambda-1){\bf r}$.  Accordingly, the microscopic expression for the displacement vector field should be taken as
\be\label{local_order_param}
\hat{\bf u}({\bf r};\Gamma) = -\, \pmb{\cal N}^{-1}\cdot \int {\rm d}{\bf r'} \, \Delta({\bf r}-{\bf r'})\, \pmb{\nabla}' n_{\rm eq}({\bf r'}) \, \left[\hat n({\bf r'};\Gamma)-n_{\rm eq}({\bf r'})\right]
\ee
with the tensor $\pmb{\cal N}=({\cal N}^{ab})$ given by
\be\label{N-dfn}
{\cal N}^{ab} \equiv \frac{1}{v} \int_{v} \nabla^a n_{\rm eq} \, \nabla^b n_{\rm eq} \, {\rm d}{\bf r} \, ,
\ee
where $v$ is the volume of a primitive unit cell of the lattice.  Since this tensor is symmetric $\pmb{\cal N}=\pmb{\cal N}^{\rm T}$, the force density field can be identified with
\be
{\bf f}({\bf r}) =-\, \pmb{\cal N}\cdot {\bf C}({\bf r}) \, .
\ee

In cubic crystals, the microscopic expression of references~\cite{SE93,S97} for the displacement vector is recovered, in which case the tensor~(\ref{N-dfn}) should be proportional to the identity tensor: ${\cal N}^{ab}={\cal N}\, \delta^{ab}$ with
\be
{\cal N} \equiv \frac{1}{3v} \int_{v} (\pmb{\nabla} n_{\rm eq})^2 \, {\rm d}{\bf r} \, .
\ee
Moreover, the equilibrium density can be expanded into lattice Fourier modes as
\be\label{n_eq-G}
n_{\rm eq}({\bf r}) = \sum_{\bf G} n_{{\rm eq},{\bf G}} \, {\rm e}^{\imath {\bf G}\cdot{\bf r}}
\ee
in terms of the vectors $\bf G$ of the reciprocal lattice.
Using the inverse Fourier transform~(\ref{IFT}) and the definition~(\ref{Delta-dfn}), we find that
\be
\hat{\bf u}({\bf r};\Gamma) = -\frac{\imath}{\cal N} \sum_{\bf G} {\bf G} \, n_{{\rm eq},{\bf G}} \int_{\cal B} \frac{{\rm d}{\bf k}}{(2\pi)^3} \, {\rm e}^{\imath{\bf k}\cdot{\bf r}}  \left[\hat{\tilde n}({\bf k}-{\bf G};\Gamma)-\tilde n_{\rm eq}({\bf k}-{\bf G})\right] ,
\ee
which is precisely the expression given in references~\cite{SE93,S97} for cubic lattices.  We note that the displacement vector field is vanishing in fluid phases where the equilibrium density is uniform and $\pmb{\nabla}n_{\rm eq}=0$, as expected if the continuous symmetry is not broken.

In Cartesian components, the local order parameter~(\ref{local_order_param}) is given by
\be\label{order_param}
\hat{u}^a({\bf r};\Gamma) = -\, (\pmb{\cal N}^{-1})^{ab} \int {\rm d}{\bf r'}  \, \Delta({\bf r}-{\bf r'})\, {\nabla'}^b n_{\rm eq}({\bf r'}) \, \left[\hat n({\bf r'};\Gamma)-n_{\rm eq}({\bf r'})\right] .
\ee
As a consequence of equation~(\ref{eq-rho}), its time evolution can be expressed as
\be
\partial_t\hat{u}^a + \hat{J}_{u^a} = 0 \label{eq-u}
\ee
in terms of the decay rate
\be\label{micro_decay_rate}
\hat{J}_{u^a}(\mathbf{r};\Gamma)   =  - \frac{1}{m} \, (\pmb{\cal N}^{-1})^{ab} \int {\rm d}{\bf r'} \, \Delta({\bf r}-{\bf r'})\, {\nabla'}^b n_{\rm eq}({\bf r'}) \, {\nabla'}^c \hat{g}^c(\mathbf{r}';\Gamma)\, .
\ee

Now, the time evolution equation of the strain tensor $\hat{u}^{ab} =(\nabla^a \hat{u}^b + \nabla^b \hat{u}^a)/2$ introduced in equation~(\ref{strain-dfn}) takes the form of the local conservation equation
\be\label{eq-strain}
\partial_t\hat{u}^{ab} + \nabla^c\hat{J}^c_{u^{ab}}  = 0 
\ee
with the associated current density
\be
\hat{J}^c_{u^{ab}}(\mathbf{r};\Gamma) = \frac{1}{2}\left(\delta^{ac}\delta^{bd}+\delta^{ad}\delta^{cb}\right)\hat{J}_{u^d}(\mathbf{r};\Gamma)\, .
\ee
Equation~(\ref{eq-strain}) along with equations~(\ref{eq-rho}), (\ref{eq-e}), and~(\ref{eq-g}) are ruling the microscopic hydrodynamics of crystals and they can be written in general form as
\be
\partial_t\, \hat{c}^{\alpha}(\mathbf{r},t)   + \nabla^a \hat{J}^{a}_{c^\alpha}(\mathbf{r},t)  = 0 \label{eqs-c}
\ee
with the following densities and current densities,
\be\label{densities}
(\hat{c}^{\alpha}) = (\hat{e},\hat{\rho},\hat{g}^b,\hat{u}^{bc}) \qquad \mbox{and} \qquad (\hat{J}^{a}_{c^\alpha}) = (\hat{J}^{a}_{e},\hat{J}^{a}_{\rho},\hat{J}^{a}_{g^b},\hat{J}^{a}_{u^{bc}}) \, .
 \ee

Since equation~(\ref{eq-strain}) is the direct consequence of equation~(\ref{eq-u}) for the displacement vector field combined with the definition~(\ref{strain-dfn}) of the strain tensor, equations~(\ref{eq-rho}), (\ref{eq-e}), (\ref{eq-g}), and~(\ref{eq-u}) form the minimal set of eight equations ruling the eight hydrodynamic modes, including the five modes resulting from the five fundamentally conserved quantities (i.e., mass, energy, and momentum) and the three additional Nambu-Goldstone modes generated by the spontaneous symmetry breaking of three-dimensional continuous spatial translations in crystals.

Accordingly, the methods developed in reference~\cite{MG20} for general continuous symmetry breaking in matter can here be applied to crystals, where three continuous symmetries are broken.  The local order parameters denoted $\hat{x}^{\alpha}$ in reference~\cite{MG20} correspond to the three components of the microscopic displacement vector $\hat{u}^a$ defined by equation~(\ref{order_param}).  The gradients $\hat{u}^{a\alpha}=\nabla^a\hat{x}^{\alpha}$ of the order parameters correspond to the symmetric strain tensor~$\hat{u}^{ab}=\hat{u}^{ba}$ defined by equation~(\ref{strain-dfn}), and the conjugated fields $\phi^{a\alpha}$ to the symmetric tensor $\phi^{ab}$ giving the force density~(\ref{force_density}).  In order to apply the results of reference~\cite{MG20} to crystals, the Greek indices $\alpha,\beta,\dots$ should thus be replaced by Latin indices $a,b,\dots$$=1,2,3=x,y,z$, and the tensors $\hat{u}^{a\alpha}$ and $\phi^{a\alpha}$ should moreover be symmetrized.


\section{Nonequilibrium statistical mechanics}
\label{Sec:NESM}

In this section, the local equilibrium approach known for normal fluids \cite{McL60,McL61,McL63,DMcL65,Z66,R66,R67,P68,DK72,OL79,BZD81,KO88,BY80,AP81,Sp91,S14} is extended to crystals using the results of reference~\cite{MG20}.  The Green-Kubo formulas are obtained for all the crystalline transport coefficients and the vacancy concentration is introduced.

\subsection{Local equilibrium distribution}

In order to deduce the macroscopic equations from the microscopic dynamics, we consider the local equilibrium distribution
 \be\label{p-leq}
p_{\rm leq}(\Gamma;\boldsymbol{\lambda}) = \frac{1}{\Delta\Gamma} \, \exp \left[ - \lambda^{\alpha}\ast\hat{c}^{\alpha} (\Gamma)- \Omega(\boldsymbol{\lambda}) \right] ,
 \ee
expressed in terms of the densities $\hat{\bf c}=(\hat{c}^{\alpha})$ defined in equation~(\ref{densities}), the conjugated fields $\boldsymbol{\lambda}=(\lambda^{\alpha})$, the integral $f\ast g \equiv \int {\rm d}\mathbf{r}\, f(\mathbf{r}) \, g(\mathbf{r})$ over the volume of the system, and the functional
 \be\label{Omega-dfn}
\Omega(\boldsymbol{\lambda}) = \ln \int\frac{{\rm d}\Gamma}{\Delta\Gamma}\, \exp\left[ - \lambda^{\alpha}\ast\hat{c}^{\alpha}(\Gamma)\right] \, ,
 \ee
 such that the local equilibrium distribution is normalized to the unit value using equation~(\ref{normalization}).  The mean values of the densities are thus given by the following functional derivatives with respect to the conjugated fields,
\be\label{c-Omega-lambda}
c^{\alpha}({\bf r}) \equiv \langle\hat{c}^{\alpha}(\mathbf{r};\Gamma)\rangle_{{\rm leq},\boldsymbol{\lambda}} =- \frac{\delta\Omega(\boldsymbol{\lambda})}{\delta\lambda^{\alpha}({\bf r})} \, . 
\ee

The  Legendre transform of the functional~(\ref{Omega-dfn}) gives the entropy functional
\be\label{entropy}
S({\bf c}) = \inf_{\boldsymbol{\lambda}}\left[\lambda^\alpha\ast {c}^{\alpha}+\Omega(\boldsymbol{\lambda}) \right] 
\ee
in units where Boltzmann's constant is equal to $k_{\rm B}=1$ \cite{McL63,AP81,S14}.
The comparison with the Euler thermodynamic relation known in crystals~\cite{MPP72,FC76} allows us to identify (up to possible corrections going as the gradient square) the conjugated fields as
\be
\lambda_{e} = \beta + O(\nabla^2) \, , \quad 
\lambda_{\rho} =-\beta\, \mu + O(\nabla^2) \, , \quad
\lambda_{g^a} =-\beta\, v^a+ O(\nabla^2) \, , \quad 
\lambda_{u^{ab}} =-\beta\, \phi^{ab}+ O(\nabla^2) \, , 
\ee
where $\beta=\beta({\bf r},t)$ is the local inverse temperature, $\mu({\bf r},t)$ the local chemical potential, $v^{a}({\bf r},t)$ the velocity field, and $\phi^{ab}({\bf r},t)$ the field introduced in equation~(\ref{Hext-phi-u}). Accordingly, the entropy~(\ref{entropy}) can be written as $S({\bf c}) = \int s({\bf c}) \, {\rm d}{\bf r} + O(\nabla^2)$ in terms of the entropy density given by Euler's relation,
\be
s = \beta(e+p)-\beta\,\mu\, \rho -\beta\, v^a g^a -\beta\, \phi^{ab} u^{ab} \, ,
\ee
and the functional~(\ref{Omega-dfn}) as $\Omega=\int\beta\, p \, {\rm d}{\bf r}+O(\nabla^2)$ with the hydrostatic pressure $p({\bf r},t)$.  The formalism is thus consistent with known local equilibrium thermodynamics in crystals.

\subsection{Time evolution}

The time evolution of any phase-space probability distribution is ruled by Liouville's equation $\partial_tp_t=-{\cal L}p_t$, where  ${\cal L}(\cdot)\equiv \{\cdot,H\}$ is the Liouvillian operator defined as the Poisson bracket with the Hamiltonian function of the microscopic dynamics.  Since the Hamiltonian function~(\ref{Hamilt_fn}) is time independent, the probability density at time $t$ is given by $p_t=\exp(-{\cal L}t) p_0$ in terms of the initial density $p_0$.  This latter is taken as a local equilibrium distribution~(\ref{p-leq}) with some initial conjugated fields $\pmb{\lambda}_0$.  However, the probability density does not keep the form~(\ref{p-leq}) of a local equilibrium distribution during the time evolution.  Nevertheless, the phase-space dynamics is point-like, so that it is possible to express the probability density at time $t$ as \cite{McL63,Z66,S14}
\be\label{p-t}
p_{t}(\Gamma) = {\rm e}^{-{\mathcal L} t} p_{\rm leq}(\Gamma;\boldsymbol{\lambda}_0) = p_{\rm leq}(\Gamma;\boldsymbol{\lambda}_t) \, {\rm e}^{\Sigma_t(\Gamma)} \, , 
\ee
by multiplying the local equilibrium distribution corresponding to time-evolved conjugated fields $\pmb{\lambda}_t$ with the exponential of the following quantity,
\be
\Sigma_t(\Gamma) \equiv   \int_0^t {\rm d}\tau\,  \partial_\tau \left[{\lambda}_\tau^\alpha\ast\hat{c}^\alpha(\Gamma_{\tau - t}) + \Omega(\boldsymbol{\lambda}_\tau)\right] . \label{Sigma}
\ee
The mean value of this quantity represents the entropy~(\ref{entropy}) that is produced during the time interval $t$
\be\label{2nd_law}
\langle \Sigma_t(\Gamma) \rangle_t = S({\bf c}_t)- S({\bf c}_0) \geq 0\, ,
 \ee
which can be shown to be always non-negative in agreement with the second law of thermodynamics \cite{S14}.  Since the system is isolated, the time derivative of the entropy is giving the entropy production rate ${\rm d}S/{\rm d}t={\rm d}_{\rm i}S/{\rm d}t\ge 0$, which can thus be calculated using equation~(\ref{Sigma}).

Now, the macroscopic local conservation equations can be obtained by averaging their microscopic analogies~(\ref{eqs-c}) over the time-evolved probability distribution~(\ref{p-t}), leading to
\be\label{bal-eq-c}
\partial_t\, c^{\alpha}  + \nabla^a\left(\bar{J}^{a}_{c^\alpha}+{\mathcal J}^{a}_{c^\alpha}\right)  = 0 \, ,
\ee
where $c^{\alpha}$ are the mean densities~(\ref{c-Omega-lambda}) at time $t$ and
\bea
&&\bar{J}^{a}_{c^{\alpha}}(\mathbf{r},t) \equiv \langle \hat{J}^a_{c^{\alpha}}(\mathbf{r};\Gamma)\rangle_{{\rm leq},\boldsymbol{\lambda}_t} \, , \label{dfn-J-bar}\\
&& {\mathcal{J}}^a_{c^{\alpha}}(\mathbf{r},t) \equiv \langle \hat{J}^a_{c^{\alpha}}(\mathbf{r};\Gamma)\big[{\rm e}^{\Sigma_t(\Gamma)}-1\big]\rangle_{{\rm leq},\boldsymbol{\lambda}_t} \label{dfn-J-cal}
\eea
are respectively the dissipativeless and dissipative current densities \cite{McL63,S14}.  This identification is justified because they satisfy the following relations,
\bea
&& \nabla^a\lambda^{\alpha}_t\ast \bar{J}^{a}_{c^{\alpha}}(t) = 0 \, , \label{eq-J-bar}\\
\frac{{\rm d}S}{{\rm d}t} &=& \nabla^a\lambda^{\alpha}_t\ast {\mathcal{J}}^a_{c^{\alpha}}(t) \ge 0 \, . \label{eq-J-cal}
\eea
Indeed, equation~(\ref{eq-J-bar}) shows that the mean values of the microscopic current densities over the local equilibrium distribution at time $t$ do not contribute to the entropy production rate.  Therefore, dissipation (i.e., entropy production) is generated according to equation~(\ref{eq-J-cal}) by the contributions~(\ref{dfn-J-cal}) to the mean values of the microscopic current densities over the full probability distribution~(\ref{p-t}) at time $t$.  In this regard, it is justified to identify equations~(\ref{dfn-J-bar}) and~(\ref{dfn-J-cal}) as the dissipativeless and dissipative current densities.  These latter are thus expected to provide the transport coefficients as Green-Kubo formulas after their expansion to first order in the gradients of the conjugated fields.

\subsection{Dissipativeless current densities}

The dissipativeless time evolution is obtained by considering the local equilibrium mean values of the densities and current densities.  

In particular, the local equilibrium mean values of the densities for momentum and energy are respectively given by
\bea
&&\langle\hat{g}^a\rangle_{\rm leq} = \rho\, v^a \, ,\label{g-leq}\\
&&\langle\hat{e}\rangle_{\rm leq} = e = e_0+\rho{\bf v}^2/2\, ,\label{e-leq}
\eea
in terms of the mean mass density $\rho=\langle\hat{\rho}\rangle_{\rm leq}=m\langle\hat{n}\rangle_{\rm leq}$ and the velocity field ${\bf v}=(v^a)$, $e_0$ denoting the internal energy in the frame moving with the crystal element.  As a consequence of~(\ref{g-leq}), equation~(\ref{eq-rho}) becomes the continuity equation $\partial_t\rho+\pmb{\nabla}\cdot(\rho{\bf v})=0$, expressing the local conservation of mass.

Furthermore, the microscopic current densities are averaged over the local equilibrium distribution~(\ref{p-leq}) to obtain the dissipativeless current densities~(\ref{dfn-J-bar}).
For the decay rate~(\ref{micro_decay_rate}) of the microscopic displacement vector~(\ref{order_param}), the local equilibrium mean value is given by
\be
\bar{J}_{u^a}({\bf r},t) \equiv \langle\hat{J}_{u^{a}}(\mathbf{r};\Gamma)\rangle_{{\rm leq},\boldsymbol{\lambda}_t} =  - v^a(\mathbf{r},t)\, ,
\label{Ju-leq}
\ee
as shown in Appendix~\ref{AppA}. 
In references~\cite{MPP72,MG20}, the local equilibrium mean value of the order parameter was supposed to have the following general form,
\be
\bar{J}_{u^c} = -A^{ac} v^a - B^{abc} \nabla^a v^b + O(\nabla^2)
\ee
with some coefficients $A^{ab}$ and $B^{abc}$, such that $\nabla^aA^{bc}=0$.  The comparison with the microscopic result~(\ref{Ju-leq}) shows that, for crystals, these coefficients are equal to
\be
A^{ab} = \delta^{ab} \qquad\mbox{and} \qquad B^{abc} =0 \, .\label{coeff-A-B}
\ee
As shown in reference~\cite{MG20}, the local equilibrium mean values of the current densities for momentum and energy are thus given by
\bea
&&\bar{J}^a_{g^b} = \langle\hat{J}^a_{g^b}\rangle_{\rm leq} = \rho\, v^b \, v^a - \sigma^{ab} +O(\nabla^2) \, , \ \qquad \label{Jg-leq}\\
&&\bar{J}^a_e = \langle\hat{J}^a_{e}\rangle_{\rm leq} = e\, v^a -\sigma^{ab} v^b +O(\nabla^2) \, , \label{Je-leq}
\eea
in terms of the reversible stress tensor
\be\label{stress}
\sigma^{ab} \equiv -p\, \delta^{ab} +\phi^{ab} \, ,
\ee
where $p$ is the hydrostatic pressure and $\phi^{ab}$ the tensorial field conjugated to the strain tensor~(\ref{strain-dfn}).

Therefore, the dissipativeless rate and current densities can be expressed by equations~(\ref{Ju-leq}), (\ref{Jg-leq}), and~(\ref{Je-leq}).

\subsection{Dissipative current densities}

In crystals, the dissipative current densities and rates are defined by ${\cal J}_{c^\alpha}^{a}=({\cal J}_q^a,{\cal J}_{g^b}^a,{\cal J}_{u^a})$ with equation~(\ref{dfn-J-cal}) and the heat current density
\be\label{heat-crnt}
\mathcal{J}^a_q \equiv \mathcal{J}^a_e - v^b \mathcal{J}^a_{g^b} -\phi^{ab} \mathcal{J}_{u^b} \, .
\ee
They can be calculated with the methods of reference~\cite{MG20}.  

At first order in the affinities or thermodynamic forces ${\cal A}_{c^\alpha}^{a}=({\cal A}_q^a,{\cal A}_{g^b}^a,{\cal A}_{u^a})$ defined by the following gradients~\cite{P67,GM84,H69,N79,Callen85},
\be
{\cal A}_q^a \equiv -\frac{1}{T^2}\, \nabla^a T \, , \qquad {\cal A}_{g^b}^a \equiv -\frac{1}{T}\, \nabla^a v^b \, , \qquad {\cal A}_{u^a} \equiv - \frac{1}{T} \, \nabla^b\phi^{ba} \, , 
\ee
the dissipative current densities and rates can be expressed as
\be
{\cal J}_{c^\alpha}^{a} = \sum_{b,\beta} {\cal L}\left(_{c^\alpha}^{a}\big\vert{_{c^\beta}^{b}}\right) \, {\cal A}_{c^\beta}^{b} \, ,
\ee
in terms of the linear response coefficients given by the following Green-Kubo formulas,
\be\label{lin-coeff}
{\cal L}\left(_{c^\alpha}^{a}\big\vert{_{c^\beta}^{b}}\right) \equiv \lim_{V\to\infty} \frac{1}{k_{\rm B}V} 
\int_0^{\infty}{\rm d}t \, \langle \delta \hat{\mathbb J}^{\prime a}_{c^\alpha}(t)\, \delta \hat{\mathbb J}^{\prime b}_{c^\beta}(0)\rangle_{\rm eq} \, ,
\ee
where
\be\label{global-crnts}
\delta\hat{\mathbb J}^{\prime a}_{c^\alpha}(t) \equiv \int_V \delta\hat{J}^{\prime a}_{c^\alpha}({\bf r},t) \, {\rm d}{\bf r} 
\ee
are the microscopic global currents defined with
\bea
\delta \hat{J}^{\prime a}_{e} & \equiv& \delta \hat{J}^a_{e} -\rho^{-1}(e_0+p)\, \delta\hat{g}^a \, ,  \label{eq:currentjprimee0GenSSB}\\
\delta \hat{J}^{\prime a}_{g^b} & \equiv& \delta\hat{J}_{g^b}^a +\left(\frac{\partial \sigma^{ab}}{\partial e_0}\right)_{\rho, \boldsymbol{\mathsf u}}  \delta \hat{e} + \left(\frac{\partial \sigma^{ab}}{\partial \rho}\right)_{e_0, \boldsymbol{\mathsf u}} \delta \hat{\rho} \, , \label{eq:currentjprimeg0GenSSB}\\
\delta\hat{J}'_{u^a} & \equiv &\delta\hat{J}_{u^a} + \rho^{-1} \, \delta\hat{g}^a\; ,\label{eq:currentjprimexGenSSB}
\eea
and $\delta\hat{X}\equiv\hat{X}-\langle\hat{X}\rangle_{{\rm leq},\boldsymbol{\lambda}_t}$.  We note that the microscopic global energy current~(\ref{eq:currentjprimee0GenSSB}) determines the heat current density~(\ref{heat-crnt}), so that we shall use the notation $c^{\alpha}=q$ in the left-hand side of Eq.~(\ref{lin-coeff}) if $c^{\alpha}=e$ in its right-hand side.

\subsection{Implications of time-reversal symmetry}

According to the symmetry $H(\Theta\Gamma)=H(\Gamma)$ of the Hamiltonian function
under the time-reversal transformation $\Theta({\bf r}_i,{\bf p}_i)=({\bf r}_i,-{\bf p}_i)$, the linear response coefficients~(\ref{lin-coeff}) should obey the Onsager-Casimir reciprocal relations~\cite{O31a,O31b,C45}
\be\label{OCRR}
{\cal L}\left(_{c^\alpha}^{a}\big\vert{_{c^\beta}^{b}}\right) = \epsilon{_{c^\alpha}^{a}}  \, \epsilon_{c^\beta}^{b} \, {\cal L}\left(_{c^\beta}^{b}\big\vert{_{c^\alpha}^{a}}\right) \, , 
\ee
where $\epsilon_{c^\alpha}^{a}=\pm 1$ if $\delta\hat{\mathbb J}_{c^\alpha}^{a}$ is even or odd under time reversal  (and there is no Einstein summation here).  The quantities $\hat{J}^a_e$, $\hat{J}_{u^a}$, and $\hat{g}^a$ are odd, while $\hat{J}^a_{g^b}$, $\hat{e}$, and $\hat{\rho}$ are even.  As a consequence, the Onsager-Casimir reciprocal relations with $\epsilon_{c^\alpha}^{a}\, \epsilon_{c^\beta}^{b}=+1$ are giving
\be\label{OCRR-even}
{\cal L}\left(_{q}^{a}\big\vert{_{q}^{b}}\right) = {\cal L}\left(_{q}^{b}\big\vert{_{q}^{a}}\right) , \quad {\cal L}\left(_{g^b}^{a}\Big\vert{_{g^d}^{c}}\right)  = {\cal L}\left(_{g^d}^{c}\Big\vert{_{g^b}^{a}}\right) , 
\quad {\cal L}\left(_{u^a}\big\vert{_{u^b}}\right) = {\cal L}\left(_{u^b}\big\vert{_{u^a}}\right) , \quad\mbox{and}\quad {\cal L}\left(_{q}^{a}\big\vert{_{u^b}}\right) = {\cal L}\left(_{u^b}\big\vert{_{q}^{a}}\right) ,
\ee
and those with $\epsilon_{c^\alpha}^{a}  \, \epsilon_{c^\beta}^{b} =-1$ lead to
\be\label{OCRR-odd}
{\cal L}\left(_{g^b}^{a}\Big\vert{_{q}^{c}}\right) = - {\cal L}\left(_{q}^{c}\Big\vert{_{g^b}^{a}}\right) \qquad\mbox{and}\qquad {\cal L}\left(_{u^a}\big\vert{_{g^c}^{b}}\right) = - {\cal L}\left(_{g^c}^{b}\big\vert{_{u^a}}\right) .
\ee
Therefore, time-reversal symmetry is significantly reducing the number of independent linear response coefficients.  Since the coefficients~(\ref{OCRR-odd}) form an antisymmetric linear response matrix, they do not contribute to entropy production~\cite{MG20}.

\subsection{Transport properties}

The independent transport coefficients are thus obtained from the microdynamics with the following Green-Kubo formulas,
 \begin{align}
 	\kappa^{ab} & \equiv \frac{1}{T^2}\, {\cal L}\left(_{q}^{a}\big\vert{_{q}^{b}}\right) = \lim_{V\rightarrow \infty} \frac{1}{k_{\rm B} T^2V}\int_0^{\infty}{\rm d}t \, \langle \delta \hat{\mathbb J}^{\prime a}_{e}(t)\, \delta \hat{\mathbb J}^{\prime b}_{e}(0)\rangle_{\text{eq}}\;, \label{kappa}\\
	\xi^{ab} &\equiv \frac{1}{T}\, {\cal L}\left(_{q}^{a}\big\vert{_{u^b}}\right) = \lim_{V\rightarrow \infty}  \frac{1}{k_{\rm B} T V}\int_0^{\infty}{\rm d}t \, \langle \delta \hat{\mathbb J}^{\prime a}_{e}(t)\,  \delta \hat{\mathbb J}'_{u^b}(0)\rangle_{\text{eq}}\;, \label{xi}\\
	\zeta^{ab} & \equiv \frac{1}{T}\, {\cal L}\left(_{u^a}\big\vert{_{u^b}}\right) = \lim_{V\rightarrow \infty}  \frac{1}{k_{\rm B} T V}\int_0^{\infty}{\rm d}t \, \langle \delta \hat{\mathbb J}^{\prime}_{u^a}(t) \, \delta \hat{\mathbb J}^{\prime}_{u^b}(0)\rangle_{\text{eq}}\;, \label{zeta}\\
	\chi^{abc} & \equiv \frac{1}{T}\, {\cal L}\left(_{q}^{a}\Big\vert{_{g^c}^{b}}\right) = \lim_{V\rightarrow \infty}  \frac{1}{k_{\rm B} T V} \int_0^{\infty}{\rm d}t \, \langle \delta \hat{\mathbb J}^{\prime a}_{e}(t)\, \delta  \hat{\mathbb J}^{\prime b}_{g^c}(0)\rangle_{\text{eq}}\;, \label{chi}\\
	\theta^{abc} & \equiv \frac{1}{T}\, {\cal L}\left(_{g^b}^{a}\big\vert{_{u^c}}\right)  = \lim_{V\rightarrow \infty}  \frac{1}{k_{\rm B} T V}\int_0^{\infty}{\rm d}t \, \langle \delta \hat{\mathbb J}^{\prime a}_{g^b}(t)\, \delta \hat{\mathbb J}^{\prime}_{u^c}(0)\rangle_{\text{eq}}\;, \label{theta}\\
	\eta^{abcd} & \equiv \frac{1}{T}\, {\cal L}\left(_{g^b}^{a}\Big\vert{_{g^d}^{c}}\right) = \lim_{V\rightarrow \infty}  \frac{1}{k_{\rm B} T V} \int_0^{\infty}{\rm d}t \, \langle \delta \hat{\mathbb J}^{\prime a}_{g^b}(t)\, \delta  \hat{\mathbb J}^{\prime c}_{g^d}(0)\rangle_{\text{eq}}\;, \label{eta}
\end{align}
expressed in terms of the microscopic global currents defined by equations~(\ref{global-crnts})-(\ref{eq:currentjprimexGenSSB}).

Because of equations~(\ref{OCRR-even}) and~(\ref{OCRR-odd}), the other linear coefficients are given by
\be
{\cal L}\left(_{u^a}\big\vert{_{q}^{b}}\right) = T \, \xi^{ba} \, , \qquad {\cal L}\left(_{g^b}^{a}\Big\vert{_{q}^{c}}\right) = - T \, \chi^{cab} \, , \qquad {\cal L}\left(_{u^a}\big\vert{_{g^c}^{b}}\right) = - T \, \theta^{bca} \, ,
\ee
and we have the following symmetries $\kappa^{ab}=\kappa^{ba}$, $\zeta^{ab}=\zeta^{ba}$, and $\eta^{abcd}=\eta^{cdab}$.  The coefficients $\kappa^{ab}$ are the heat conductivities, $\eta^{abcd}$ the viscosities, and $\zeta^{ab}$ are strain friction coefficients.

Consequently, the dissipative current densities and rates take the following forms \footnote{We note that the coefficient $\chi^{abc}$ in equation~(6.19) of reference \cite{MG20} should correctly read $\chi^{cab}$.},
\bea
\mathcal{J}^a_{q} &=& -\kappa^{ab}\, \nabla^bT  -\chi^{abc}\, \nabla^b v^c - \xi^{ac}\, \nabla^b\phi^{bc}\, ,  \label{dissip-J_e}\\
\mathcal{J}^a_{g^b} &=& \frac{\chi^{cab}}{T} \, \nabla^c T  -\eta^{abcd} \, \nabla^c v^d- \theta^{abd}\nabla^c\phi^{cd}\, , \label{dissip-J_g^b}\\
\mathcal{J}_{u^a} &=& -\frac{\xi^{ba}}{T} \, \nabla^bT  +\theta^{bca}\, \nabla^b v^c - \zeta^{ac}\, \nabla^b\phi^{bc}\, . \label{dissip-J_u}
\eea

\subsection{Implications of crystallographic symmetries}

According to Curie's principle \cite{C1894}, the tensorial properties should be symmetric under the transformations of the crystallographic group.  This principle applies, in particular, to the rank-three tensors~(\ref{chi}) and~(\ref{theta}) describing cross effects coupling the transport of momentum to those of heat and crystalline order.  In isotropic phases, rank-three tensors are always vanishing, because such phases have symmetry centers.  However, this is no longer the case in anisotropic phases, as illustrated with the phenomenon of piezoelectricity, which is also described by a rank-three tensor~\cite{LLv8}.  Among the 32 possible crystallographic point groups (also called classes), 20 of them are known to allow for non-vanishing rank-three tensors:
\bea
&&{\rm C}_1\, , \ {\rm C}_{\rm s} \, , \ {\rm C}_2 \, , \ {\rm C}_{2{\rm v}} \, , \ {\rm C}_3 \, , \ {\rm C}_{3{\rm v}} \, , \ {\rm C}_4 \, , \ {\rm C}_{4{\rm v}} \, , \ {\rm C}_6 \, , \ {\rm C}_{6{\rm v}} \, , \label{pyro} \\
&&{\rm D}_2\, , \ {\rm D}_{2{\rm d}} \, , {\rm C}_{3{\rm h}} \, , \ {\rm D}_3 \, , \ {\rm D}_{3{\rm h}} \, , \ {\rm D}_4 \, , \ {\rm S}_4 \, , \ {\rm D}_{6} \, , \ {\rm T} \, , \ {\rm T}_{\rm d} \, . \label{piezo}
\eea
The 10 classes~(\ref{pyro}) are compatible with pyroelectricity and the 20 classes~(\ref{pyro}) and~(\ref{piezo}) with piezoelectrictity~\cite{LLv8}.  Rank-three tensors are vanishing in the 12 other classes because these point groups are either centrosymmetric (i.e., they contain the inversion ${\bf r}\to -{\bf r}$ with respect to a symmetry center), or they contain $90^{\circ}$ rotations around axes perpendicular to the faces as for the cubic (or orthohedral) crystallographic group O.  Therefore, the transport coefficients~(\ref{chi}) and~(\ref{theta}) may be non-vanishing in the 20 crystallographic classes~(\ref{pyro}) and~(\ref{piezo}).

\subsection{Vacancy concentration}

In the crystal at equilibrium, the macrofield of particle density is uniform and equal to the component ${\bf G}=0$ of the lattice Fourier expansion~(\ref{n_eq-G}) and, equivalently, to the macrofield~(\ref{macrofield-condition}) corresponding to the periodic equilibrium density $n_{\rm eq}({\bf r})$:
\be\label{n_eq-0}
n_{{\rm eq},0} = \frac{1}{v} \int_v n_{\rm eq}({\bf r})\, {\rm d}{\bf r} = \int \Delta({\bf r}-{\bf r'}) \, n_{\rm eq}({\bf r'}) \, {\rm d}{\bf r'} \, .
\ee
Under nonequilibrium conditions, the macrofield of particle density may deviate with respect to its equilibrium value by two possible mechanisms: (1) the lattice dilatation or contraction corresponding to the strain $\pmb{\nabla}\cdot{\bf u}=\nabla^a u^a$; (2) vacancies or interstitials, decreasing or increasing the occupancy of the lattice cells by particles \cite{MPP72,FC76,SE93,S97}.  To describe these latter, a macrofield giving the density of vacancies, also called vacancy concentration, can be defined as
\be\label{vacancy_conc}
\hat{c}({\bf r};\Gamma) \equiv - \int \Delta({\bf r}-{\bf r'}) \left[ \hat{n}({\bf r'};\Gamma)- n_{{\rm eq},0} + n_{{\rm eq},0} \, \nabla^a \hat{u}^a({\bf r'};\Gamma)\right] {\rm d}{\bf r'} \, .
\ee
The time evolution of this macrofield is driven by the local conservation equation~(\ref{eq-rho}) for the mass density $\hat\rho=m\hat n$ and by equation~(\ref{eq-u}) for the displacement vector~(\ref{order_param}).

The local equilibrium mean value of the vacancy concentration can be expressed as
\be
c({\bf r},t) \equiv \langle\hat{c}({\bf r};\Gamma)\rangle_{{\rm leq},\pmb{\lambda}_t} = - \delta n({\bf r},t) - n_{{\rm eq},0} \, \nabla^a u^a({\bf r},t) \, , 
\ee
where $\delta n({\bf r},t)$ denotes the macrofield giving the deviation of the mean particle density with respect to its equilibrium value~(\ref{n_eq-0}).

Accordingly, the dynamics of the vacancy concentration~(\ref{vacancy_conc}) is driven by the evolution equations~(\ref{eq-rho}) and~(\ref{eq-u}) for the particle density $\hat{n}=\hat{\rho}/m$ and the displacement vector $\hat{u}^a$.

\section{Macroscopic description}
\label{Sec:Macro}

Here, the eight hydrodynamic modes of the crystal are deduced from the linearized macroscopic equations and their dispersion relations are obtained by using expansions in powers of the gradients with respect to the dissipativeless elastic dynamics.

\subsection{Macroscopic equations}

As shown in the previous section~\ref{Sec:NESM}, the macroscopic equations ruling the hydrodynamics of crystals can be obtained from the microscopic local conservation equations~(\ref{eq-rho}), (\ref{eq-e}), and~(\ref{eq-g}) for mass, energy, and momentum, combined with the evolution equation~(\ref{eq-u}) for the microscopic displacement vector~(\ref{order_param}), using the local equilibrium distribution~(\ref{p-leq}) and systematic expansions in powers of the gradients.  The transport coefficients are thus given by the Green-Kubo formulas~(\ref{kappa})-(\ref{eta}), entering the expressions of the dissipative current densities~(\ref{dissip-J_e})-(\ref{dissip-J_u}). Gathering them with the dissipativeless current densities~(\ref{g-leq}), (\ref{Ju-leq}), (\ref{Jg-leq}), and~(\ref{Je-leq}) (respectively, for mass, displacement, momentum, and energy) and the heat current density~(\ref{heat-crnt}), the following macroscopic equations are obtained,
\bea
\partial_t \rho  +\nabla^a(\rho v^a) &=& 0 \, ,\\
\partial_t e+\nabla^a( e\, v^a - \sigma^{ab} v^b + v^b \mathcal{J}^a_{g^b} +\phi^{ab} \mathcal{J}_{u^b} + {\mathcal J}^a_q) &=& 0 \, ,\\
\partial_t (\rho v^b) +\nabla^a( \rho v^a v^b -\sigma^{ab}  + {\mathcal J}^a_{g^b}) &=& 0 \, ,\\
\partial_t u^a -v^a + {\mathcal J}_{u^a}  &=& 0 \, , 
 \eea
with the stress tensor~(\ref{stress}) and the dissipative current densities~(\ref{dissip-J_e})-(\ref{dissip-J_u}).  

\subsection{Linearized macroscopic equations}

In order to investigate the time evolution of small deviations in the macrofields around equilibrium, 
the macroscopic equations are linearized, leading to the following forms,
 \bea
&& \partial_t \rho = - \rho \, \nabla^a v^a \, , \label{macro-eq-rho}\\
&& \partial_t e_0 = -(e_0+p) \nabla^a v^a + \kappa^{ab} \nabla^a\nabla^b T + \chi^{abc} \nabla^a\nabla^b v^c + \xi^{ac} \nabla^a\nabla^b\phi^{bc} \, , \label{macro-eq-e0}\\
&& \rho\, \partial_t v^b = - \nabla^b p + \nabla^a \phi^{ab} +\theta^{abd} \nabla^a\nabla^c \phi^{cd}  - \frac{\chi^{cab}}{T} \, \nabla^a\nabla^c T + \eta^{abcd} \nabla^a\nabla^c v^d \, , \label{macro-eq-v}\\
&& \partial_t u^a = v^a -\theta^{bca} \nabla^b v^c +\frac{\xi^{ba}}{T} \, \nabla^b T + \zeta^{ac} \nabla^b \phi^{bc} \, , \label{macro-eq-u}
\eea
respectively, for the mean mass density $\rho$, the mean internal energy $e_0$, the velocity field $v^b$, and the mean displacement vector $u^a$.

The mass density macrofield can be decomposed as
\be
\rho=m \left( n_{{\rm eq},0}- n_{{\rm eq},0} \, \nabla^a u^a-c\right)
\ee
in terms of the mean equilibrium density $n_{{\rm eq},0}$ of the atoms of mass $m$, the contribution from the trace of the strain tensor $u^{aa}=\nabla^a u^a$, and the vacancy concentration $c$.  Introducing the fraction of vacancies as
\be\label{vacancy_fraction}
{\mathfrak y} \equiv \frac{c}{n_{{\rm eq},0}} \, ,
\ee
the mass density $\rho$ is thus determined by the trace of the strain tensor $u^{aa}=\nabla^a u^a$ and the vacancy fraction~${\mathfrak y}$.  In particular, the deviations of the mass density with respect to the mean equilibrium mass density $\rho\simeq\rho_{{\rm eq},0}=m n_{{\rm eq},0}$ can be written as
\be
\delta\rho = - \rho \left( \nabla^a\delta u^a + \delta{\mathfrak y} \right) ,
\ee
where $\delta$ stands for either $\partial_t$ or $\pmb{\nabla}$.  Accordingly, equation~(\ref{macro-eq-rho}) can be replaced by the evolution equation for the vacancy fraction by using equation~(\ref{macro-eq-u}) for the displacement vector $u^a$:
 \be
\partial_t {\mathfrak y} = \nabla^a v^a - \nabla^a\partial_t u^a = \theta^{bca} \nabla^a\nabla^b v^c -\frac{\xi^{ba}}{T} \, \nabla^a\nabla^b T - \zeta^{ac} \nabla^a\nabla^b \phi^{bc} \, . \label{macro-eq-y}
\ee

Furthermore, we may introduce the entropy per unit mass ${\mathfrak s}\equiv S/M$ in every element of the crystal, which is linked to the energy density $e_0$, the mass density $\rho$, and the pressure $p$ by the following Gibbs relation
\be
T\rho \, \delta{\mathfrak s} = \delta e_0 - \frac{e_0+p}{\rho} \, \delta\rho
\ee
for small deviations with respect to equilibrium.  Using equations~(\ref{macro-eq-rho}) and~(\ref{macro-eq-e0}), the evolution equation for the entropy per unit mass is thus given by
\be
T\rho \, \partial_t{\mathfrak s} = \kappa^{ab} \nabla^a\nabla^b T + \chi^{abc} \nabla^a\nabla^b v^c + \xi^{ac} \nabla^a\nabla^b\phi^{bc} \, . \label{macro-eq-s}
\ee

Besides, the velocity field obeys equation~(\ref{macro-eq-v}) and the time evolution of the strain tensor $u^{ab}=(\nabla^a u^b+\nabla^b u^a)/2$ can be deduced from equation~(\ref{macro-eq-u}).

In order to close the set of equations~(\ref{macro-eq-v}), (\ref{macro-eq-u}), (\ref{macro-eq-y}), and~(\ref{macro-eq-s}), we use
\bea
\delta T &=& \left(\frac{\partial T}{\partial{\mathfrak y}}\right)_{{\mathfrak s},\boldsymbol{\mathsf u}} \delta{\mathfrak y} + \left(\frac{\partial T}{\partial{\mathfrak s}}\right)_{{\mathfrak y},\boldsymbol{\mathsf u}} \delta{\mathfrak s} + \left(\frac{\partial T}{\partial u^{ab}}\right)_{{\mathfrak s},{\mathfrak y}} \delta u^{ab} \, , \\
\delta p &=& \left(\frac{\partial p}{\partial{\mathfrak y}}\right)_{{\mathfrak s},\boldsymbol{\mathsf u}} \delta{\mathfrak y} + \left(\frac{\partial p}{\partial{\mathfrak s}}\right)_{{\mathfrak y},\boldsymbol{\mathsf u}} \delta{\mathfrak s} + \left(\frac{\partial p}{\partial u^{ab}}\right)_{{\mathfrak s},{\mathfrak y}} \delta u^{ab} \, , \\
\delta \phi^{ab} &=& \left(\frac{\partial \phi^{ab} }{\partial{\mathfrak y}}\right)_{{\mathfrak s},\boldsymbol{\mathsf u}} \delta{\mathfrak y} + \left(\frac{\partial \phi^{ab} }{\partial{\mathfrak s}}\right)_{{\mathfrak y},\boldsymbol{\mathsf u}} \delta{\mathfrak s} + \left(\frac{\partial \phi^{ab} }{\partial u^{cd}}\right)_{{\mathfrak s},{\mathfrak y}} \delta u^{cd} \, , 
\eea
and, as a consequence, a similar expansion for the stress tensor~(\ref{stress}).

Accordingly, the linearized evolution equation~(\ref{macro-eq-y}) for the vacancy fraction~(\ref{vacancy_fraction}) is transformed into
 \be\label{eq-dy}
\partial_t {\mathfrak y} = D_{\mathfrak y}^{ab} \nabla^a\nabla^b{\mathfrak y} + D_{{\mathfrak y}{\mathfrak s}}^{ab} \nabla^a\nabla^b{\mathfrak s} + F_{{\mathfrak y}{\bf v}}^{abc} \nabla^a\nabla^b v^c + D_{{\mathfrak y}{\bf u}}^{abcd} \nabla^a\nabla^b\nabla^c u^d
\ee
with the following coefficients,
\bea
&& D_{\mathfrak y}^{ab} \equiv -\frac{\xi^{ba}}{T}\, \left(\frac{\partial T}{\partial{\mathfrak y}}\right)_{{\mathfrak s},\boldsymbol{\mathsf u}} -\zeta^{ac} \left(\frac{\partial \phi^{bc} }{\partial{\mathfrak y}}\right)_{{\mathfrak s},\boldsymbol{\mathsf u}} \, , \\
&& D_{{\mathfrak y}{\mathfrak s}}^{ab} \equiv -\frac{\xi^{ba}}{T}\, \left(\frac{\partial T}{\partial{\mathfrak s}}\right)_{{\mathfrak y},\boldsymbol{\mathsf u}} -\zeta^{ac} \left(\frac{\partial \phi^{bc} }{\partial{\mathfrak s}}\right)_{{\mathfrak y},\boldsymbol{\mathsf u}}\, , \\
&& D_{{\mathfrak y}{\bf u}}^{abcd} \equiv -\frac{\xi^{ba}}{T}\, \left(\frac{\partial T}{\partial u^{cd}}\right)_{{\mathfrak s},{\mathfrak y}} -\zeta^{ae} \left(\frac{\partial \phi^{be} }{\partial u^{cd}}\right)_{{\mathfrak s},{\mathfrak y}} \, , \\
&& F_{{\mathfrak y}{\bf v}}^{abc} \equiv \theta^{bca} \, .
\eea

Equation~(\ref{macro-eq-s}) for the evolution of the entropy per unit mass takes the form
\be\label{eq-ds}
\partial_t {\mathfrak s} = D_{{\mathfrak s}{\mathfrak y}}^{ab} \nabla^a\nabla^b{\mathfrak y} + D_{\mathfrak s}^{ab} \nabla^a\nabla^b{\mathfrak s} + F_{{\mathfrak s}{\bf v}}^{abc} \nabla^a\nabla^b v^c + D_{{\mathfrak s}{\bf u}}^{abcd} \nabla^a\nabla^b\nabla^c u^d
\ee
with
\bea
&& D_{{\mathfrak s}{\mathfrak y}}^{ab} \equiv \frac{\kappa^{ab}}{T\rho} \left(\frac{\partial T}{\partial{\mathfrak y}}\right)_{{\mathfrak s},\boldsymbol{\mathsf u}} + \frac{\xi^{ac}}{T\rho} \left(\frac{\partial \phi^{bc} }{\partial{\mathfrak y}}\right)_{{\mathfrak s},\boldsymbol{\mathsf u}}\, , \\
&& D_{\mathfrak s}^{ab} \equiv \frac{\kappa^{ab}}{T\rho} \left(\frac{\partial T}{\partial{\mathfrak s}}\right)_{{\mathfrak y},\boldsymbol{\mathsf u}} + \frac{\xi^{ac}}{T\rho} \left(\frac{\partial \phi^{bc} }{\partial{\mathfrak s}}\right)_{{\mathfrak y},\boldsymbol{\mathsf u}}\, , \\
&& D_{{\mathfrak s}{\bf u}}^{abcd} \equiv \frac{\kappa^{ab}}{T\rho} \left(\frac{\partial T}{\partial u^{cd}}\right)_{{\mathfrak s},{\mathfrak y}} + \frac{\xi^{ae}}{T\rho}  \left(\frac{\partial \phi^{be} }{\partial u^{cd}}\right)_{{\mathfrak s},{\mathfrak y}}\, , \\
&& F_{{\mathfrak s}{\bf v}}^{abc} \equiv \frac{\chi^{abc}}{T\rho} \, .
\eea

Equation~(\ref{macro-eq-u}) for the displacement vector becomes
\be\label{eq-du}
\partial_t u^b = v^b + F_{{\bf u}{\bf v}}^{abc} \nabla^a v^c + D_{{\bf u}{\mathfrak y}}^{ab} \nabla^a{\mathfrak y} + D_{{\bf u}{\mathfrak s}}^{ab} \nabla^a{\mathfrak s} + D_{\bf u}^{abcd} \nabla^a\nabla^c u^d
\ee
with
\bea
&& D_{{\bf u}{\mathfrak y}}^{ab} \equiv \frac{\xi^{ba}}{T}\, \left(\frac{\partial T}{\partial{\mathfrak y}}\right)_{{\mathfrak s},\boldsymbol{\mathsf u}} +\zeta^{bc} \left(\frac{\partial \phi^{ac} }{\partial{\mathfrak y}}\right)_{{\mathfrak s},\boldsymbol{\mathsf u}} = - D_{\mathfrak y}^{ba}\, , \\
&& D_{{\bf u}{\mathfrak s}}^{ab} \equiv \frac{\xi^{ba}}{T}\, \left(\frac{\partial T}{\partial{\mathfrak s}}\right)_{{\mathfrak y},\boldsymbol{\mathsf u}} +\zeta^{bc} \left(\frac{\partial \phi^{ac} }{\partial{\mathfrak s}}\right)_{{\mathfrak y},\boldsymbol{\mathsf u}} = - D_{{\mathfrak y}{\mathfrak s}}^{ba}\, , \\
&& D_{\bf u}^{abcd} \equiv \frac{\xi^{ba}}{T}\, \left(\frac{\partial T}{\partial u^{cd}}\right)_{{\mathfrak s},{\mathfrak y}} +\zeta^{be} \left(\frac{\partial \phi^{ae} }{\partial u^{cd}}\right)_{{\mathfrak s},{\mathfrak y}} = - D_{{\mathfrak y}{\bf u}}^{bacd} \, , \\
&& F_{{\bf u}{\bf v}}^{abc} \equiv - \theta^{acb}= - F_{{\mathfrak y}{\bf v}}^{bac}\, .
\eea

Finally, equation~(\ref{macro-eq-v}) for the velocity field gives
\bea
\partial_t v^b &=& C_{{\bf v}{\bf u}}^{abcd} \nabla^a\nabla^cu^d + C_{{\bf v}{\mathfrak y}}^{ab} \nabla^a{\mathfrak y} + C_{{\bf v}{\mathfrak s}}^{ab} \nabla^a{\mathfrak s} + D_{\bf v}^{abcd} \nabla^a\nabla^cv^d \nonumber\\
&& +F_{{\bf v}{\bf u}}^{abcde}\nabla^a\nabla^c\nabla^d u^e + F_{{\bf v}{\mathfrak y}}^{abc} \nabla^a\nabla^c{\mathfrak y} + F_{{\bf v}{\mathfrak s}}^{abc} \nabla^a\nabla^c{\mathfrak s} 
\label{eq-dv}
\eea
with
\bea
&& C_{{\bf v}{\bf u}}^{abcd} \equiv \frac{1}{\rho} \left(\frac{\partial \sigma^{ab} }{\partial u^{cd}}\right)_{{\mathfrak s},{\mathfrak y}} = -\frac{1}{\rho} \left(\frac{\partial p}{\partial u^{cd}}\right)_{{\mathfrak s},{\mathfrak y}} \delta^{ab} +\frac{1}{\rho} \left(\frac{\partial \phi^{ab} }{\partial u^{cd}}\right)_{{\mathfrak s},{\mathfrak y}} \, , \label{Cvu}\\
&& C_{{\bf v}{\mathfrak y}}^{ab} \equiv \frac{1}{\rho} \left(\frac{\partial \sigma^{ab} }{\partial {\mathfrak y}}\right)_{{\mathfrak s},\boldsymbol{\mathsf u}} = -\frac{1}{\rho} \left(\frac{\partial p}{\partial {\mathfrak y}}\right)_{{\mathfrak s},\boldsymbol{\mathsf u}}  \delta^{ab} +\frac{1}{\rho} \left(\frac{\partial \phi^{ab} }{\partial {\mathfrak y}}\right)_{{\mathfrak s},\boldsymbol{\mathsf u}}  \, , \\
&& C_{{\bf v}{\mathfrak s}}^{ab} \equiv \frac{1}{\rho} \left(\frac{\partial \sigma^{ab} }{\partial {\mathfrak s}}\right)_{{\mathfrak y},\boldsymbol{\mathsf u}} = -\frac{1}{\rho} \left(\frac{\partial p}{\partial {\mathfrak s}}\right)_{{\mathfrak y},\boldsymbol{\mathsf u}}  \delta^{ab} +\frac{1}{\rho} \left(\frac{\partial \phi^{ab} }{\partial {\mathfrak s}}\right)_{{\mathfrak y},\boldsymbol{\mathsf u}}  \, , \\
&& D_{\bf v}^{abcd} \equiv \frac{\eta^{abcd}}{\rho} \, , \label{diff-visco}\\
&& F_{{\bf v}{\bf u}}^{abcde} \equiv -\frac{\chi^{cab}}{T\rho} \left(\frac{\partial T}{\partial u^{de}}\right)_{{\mathfrak s},{\mathfrak y}} +\frac{\theta^{abf}}{\rho} \left(\frac{\partial \phi^{cf} }{\partial u^{de}}\right)_{{\mathfrak s},{\mathfrak y}} \, , \\
&& F_{{\bf v}{\mathfrak y}}^{abc} \equiv -\frac{\chi^{cab}}{T\rho} \left(\frac{\partial T}{\partial {\mathfrak y}}\right)_{{\mathfrak s},\boldsymbol{\mathsf u}} +\frac{\theta^{abd}}{\rho}  \left(\frac{\partial \phi^{cd} }{\partial {\mathfrak y}}\right)_{{\mathfrak s},\boldsymbol{\mathsf u}}  \, , \\
&& F_{{\bf v}{\mathfrak s}}^{abc} \equiv  -\frac{\chi^{cab}}{T\rho} \left(\frac{\partial T}{\partial {\mathfrak s}}\right)_{{\mathfrak y},\boldsymbol{\mathsf u}} +\frac{\theta^{abd}}{\rho}  \left(\frac{\partial \phi^{cd} }{\partial {\mathfrak s}}\right)_{{\mathfrak y},\boldsymbol{\mathsf u}}  \, ,
\eea
where (\ref{Cvu}) is the rank-four elasticity tensor divided by the mean mass density $\rho=m n_{{\rm eq},0}$.

The coefficients denoted with the letter $C$ are conservative (i.e., adiabatic or dissipativeless) properties, those with the letter $D$ are dissipative properties, and those with the letter $F$ are conservative coupling properties.

We note that the evolution equations~(\ref{eq-dy}), (\ref{eq-ds}), (\ref{eq-du}), and~(\ref{eq-dv}) form a closed set of linear partial differential equations, which can thus be written in matrix form as
\be\label{eq-psi}
\partial_t \, \pmb{\psi} = \hat{\boldsymbol{\mathsf L}}\cdot\pmb{\psi}  \qquad \mbox{with} \qquad
\pmb{\psi}=(\delta{\mathfrak y},\delta{\mathfrak s},\delta v^b,\delta u^b)^{\rm T} \, ,
\ee
where $\hat{\boldsymbol{\mathsf L}}$ is a $8\times 8$ matrix with elements involving the gradient operator $\pmb{\nabla}$ at the first, second, or third power, and acting on the eight-dimensional vector field $\pmb{\psi}({\bf r},t)$, the superscript T denoting the transpose.

\subsection{Hydrodynamic modes}

Supposing that the solution of equation~(\ref{eq-psi}) has the form $\pmb{\psi}({\bf r},t)\sim\exp(\imath{\bf q}\cdot{\bf r}+ zt)$, the dispersion relations $z({\bf q})$ of the eight hydrodynamic modes are provided by solving the eigenvalue problem:
\be
{\boldsymbol{\mathsf L}}\cdot\pmb{\psi} = z\, \pmb{\psi}
\ee
with the $8\times 8$ matrix
\be
{\boldsymbol{\mathsf L}}=\left(
\begin{array}{llll}
-D_{\mathfrak y}^{cd} q^c q^d  & 
-D_{{\mathfrak y}{\mathfrak s}}^{cd} q^c q^d & 
- F_{{\mathfrak y}{\bf v}}^{cdb} q^c q^d & 
-\imath D_{{\mathfrak y}{\bf u}}^{cdeb} q^c q^d q^e\\
-D_{{\mathfrak s}{\mathfrak y}}^{cd} q^c q^d  & 
-D_{\mathfrak s}^{cd} q^c q^d  & 
-F_{{\mathfrak s}{\bf v}}^{cdb} q^c q^d  &  
-\imath D_{{\mathfrak s}{\bf u}}^{cdeb} q^c q^d q^e\\
\imath \, C_{{\bf v}{\mathfrak y}}^{ca} q^c - F_{{\bf v}{\mathfrak y}}^{cad} q^c q^d &
\imath \, C_{{\bf v}{\mathfrak s}}^{ca} q^c - F_{{\bf v}{\mathfrak s}}^{cad} q^c q^d & 
-D_{\bf v}^{cadb} q^c q^d & 
-C_{{\bf v}{\bf u}}^{cadb} q^c q^d -\imath F_{{\bf v}{\bf u}}^{cadeb} q^c q^d q^e\\
\imath D_{{\bf u}{\mathfrak y}}^{ca} q^c &
\imath D_{{\bf u}{\mathfrak s}}^{ca} q^c &
\delta^{ab} + \imath F_{{\bf u}{\bf v}}^{cab} q^c &
-D_{\bf u}^{cadb} q^c q^d 
\end{array}
\right) .
\label{L}
\ee

The eigenvalue problem can be solved perturbatively starting from the dispersion relations of the dissipativeless crystal with vanishing coefficients $D$'s and $F$'s.  Such a method corresponds to expanding the dispersion relations in powers of the wave number $\bf q$.  For the dissipativeless crystal, the dispersion relations are linear in the wave number $q=\Vert{\bf q}\Vert$.  The next order of the perturbation calculation gives the terms of $O(q^2)$, providing the damping rates of the modes.  Since the dissipativeless and dissipative current densities are known at leading order in the gradient expansion of the fields, the corrections of $O(q^3)$ are not relevant to the calculation.

Accordingly, we consider the following expansions of the matrix~(\ref{L}), its eigenvalues, and its eigenvectors:
\be
{\boldsymbol{\mathsf L}} = {\boldsymbol{\mathsf L}}^{(0)} + {\boldsymbol{\mathsf L}}^{(1)} \, , \qquad 
\pmb{\psi} = \pmb{\psi}^{(0)} +\pmb{\psi}^{(1)} + \cdots \, , \qquad
z= z^{(0)} + z^{(1)} + \cdots
\ee
with
\be
{\boldsymbol{\mathsf L}}^{(0)}=\left(
\begin{array}{llll}
0  & 
0 & 
0 & 
0 \\
0 & 
0 & 
0 &  
0 \\
\imath \, C_{{\bf v}{\mathfrak y}}^{ca} q^c &
\imath \, C_{{\bf v}{\mathfrak s}}^{ca} q^c & 
0 & 
-C_{{\bf v}{\bf u}}^{cadb} q^c q^d \\
0 &
0 &
\delta^{ab}  &
0
\end{array}
\right)
\label{L0}
\ee
and
\be
{\boldsymbol{\mathsf L}}^{(1)}=\left(
\begin{array}{llll}
-D_{\mathfrak y}^{cd} q^c q^d  & 
-D_{{\mathfrak y}{\mathfrak s}}^{cd} q^c q^d & 
- F_{{\mathfrak y}{\bf v}}^{cdb} q^c q^d & 
-\imath D_{{\mathfrak y}{\bf u}}^{cdeb} q^c q^d q^e\\
-D_{{\mathfrak s}{\mathfrak y}}^{cd} q^c q^d  & 
-D_{\mathfrak s}^{cd} q^c q^d  & 
-F_{{\mathfrak s}{\bf v}}^{cdb} q^c q^d  &  
-\imath D_{{\mathfrak s}{\bf u}}^{cdeb} q^c q^d q^e\\
 - F_{{\bf v}{\mathfrak y}}^{cad} q^c q^d &
- F_{{\bf v}{\mathfrak s}}^{cad} q^c q^d & 
-D_{\bf v}^{cadb} q^c q^d & 
-\imath F_{{\bf v}{\bf u}}^{cadeb} q^c q^d q^e\\
\imath D_{{\bf u}{\mathfrak y}}^{ca} q^c &
\imath D_{{\bf u}{\mathfrak s}}^{ca} q^c &
\imath F_{{\bf u}{\bf v}}^{cab} q^c &
-D_{\bf u}^{cadb} q^c q^d 
\end{array}
\right) .
\label{L1}
\ee

In order to solve the eigenvalue problem at zeroth order,  we introduce the $3\times 3$ real symmetric matrix $\boldsymbol{\mathsf M}$ defined with the elements
\be
M^{ab} \equiv C_{{\bf v}{\bf u}}^{cadb} q^c q^d
\ee
and the three-dimensional vectors ${\bf N}_{\mathfrak y}$ and ${\bf N}_{\mathfrak s}$ with the components
\be
N_{\mathfrak y}^{a} \equiv C_{{\bf v}{\mathfrak y}}^{ca} q^c  \qquad\mbox{and}\qquad
N_{\mathfrak s}^{a} \equiv C_{{\bf v}{\mathfrak s}}^{ca} q^c \, .
\ee
Since the matrix $\boldsymbol{\mathsf M}$ is real symmetric, it can be diagonalized by an orthogonal transformation giving the three eigenvalues $\lambda_{\sigma}$ with $\sigma=1,2,3$, and the three eigenvectors $\pmb{\epsilon}_{\sigma}$, which are supposed to form an orthonormal basis, $\pmb{\epsilon}_{\sigma}\cdot\pmb{\epsilon}_{\sigma'}=\delta_{\sigma\sigma'}$ \cite{AM76}:
\be
\boldsymbol{\mathsf M} \cdot \pmb{\epsilon}_{\sigma} = \lambda_{\sigma} \, \pmb{\epsilon}_{\sigma} \qquad\mbox{for}\qquad \sigma=1,2,3 \, .  
\ee
Since $\boldsymbol{\mathsf M}=O(q^2)$, its eigenvalues are also going as $\lambda_{\sigma}=O(q^2)$.  The eigenvectors $\pmb{\epsilon}_{\sigma}$ also depend on the wave vector ${\bf q}$, but since they are normalized to the unit value, they only depend on the direction of the wave vector, ${\bf q}/q$.

At zeroth order, the right eigenvectors $\pmb{\psi}^{(0)}_{\alpha}$ and the left eigenvectors $\pmb{\tilde\psi}^{(0)}_{\alpha}$ associated with the eigenvalues $z^{(0)}_{\alpha}$ are obtained by solving the following eigenvalue problem,
\be
{\boldsymbol{\mathsf L}}^{(0)}\cdot\pmb{\psi}^{(0)}_{\alpha} = z^{(0)}_{\alpha}\, \pmb{\psi}^{(0)}_{\alpha} \, , \qquad {\boldsymbol{\mathsf L}}^{(0)\dagger}\cdot\pmb{\tilde\psi}^{(0)}_{\alpha} = z^{(0)*}_{\alpha}\, \pmb{\tilde\psi}^{(0)}_{\alpha} \, ,
\ee
and they are taken to satisfy the biorthonormality conditions, $\pmb{\tilde\psi}^{(0)\dagger}_{\alpha}\cdot\pmb{\psi}^{(0)}_{\beta}=\delta_{\alpha\beta}$ for $\alpha,\beta=1,2,\dots,8$.  They are given by
\bea
&& z^{(0)}_{\alpha} = -\imath \lambda_{\alpha}^{1/2} \, , \qquad \
\pmb{\psi}^{(0)}_{\alpha} =
\left(
\begin{array}{c}
0 \\
0 \\
-\imath \lambda_{\alpha}^{1/2} \pmb{\epsilon}_{\alpha} \\
\pmb{\epsilon}_{\alpha} 
\end{array}
\right) , \qquad\ \ 
\pmb{\tilde\psi}^{(0)}_{\alpha} =\frac{1}{2}
\left(
\begin{array}{c}
\imath \lambda_{\alpha}^{-1} {\bf N}_{\mathfrak y}\cdot\pmb{\epsilon}_{\alpha} \\
\imath \lambda_{\alpha}^{-1} {\bf N}_{\mathfrak s}\cdot\pmb{\epsilon}_{\alpha} \\
-\imath\lambda_{\alpha}^{-1/2} \pmb{\epsilon}_{\alpha} \\
\pmb{\epsilon}_{\alpha} 
\end{array}
\right) \quad\ (\alpha=1,2,3)\, ; 
\\
&& z^{(0)}_{\alpha+3} = +\imath \lambda_{\alpha}^{1/2} \, , \quad \ \ 
\pmb{\psi}^{(0)}_{\alpha+3} =
\left(
\begin{array}{c}
0 \\
0 \\
+\imath \lambda_{\alpha}^{1/2} \pmb{\epsilon}_{\alpha} \\
\pmb{\epsilon}_{\alpha} 
\end{array}
\right) , \qquad\ 
\pmb{\tilde\psi}^{(0)}_{\alpha+3} =\frac{1}{2}
\left(
\begin{array}{c}
\imath \lambda_{\alpha}^{-1} {\bf N}_{\mathfrak y}\cdot\pmb{\epsilon}_{\alpha} \\
\imath \lambda_{\alpha}^{-1} {\bf N}_{\mathfrak s}\cdot\pmb{\epsilon}_{\alpha} \\
+\imath\lambda_{\alpha}^{-1/2} \pmb{\epsilon}_{\alpha} \\
\pmb{\epsilon}_{\alpha} 
\end{array}
\right) \quad (\alpha=1,2,3)\, ;  \qquad
\\
&& z^{(0)}_{7} = 0 \, , \qquad\qquad\ \ 
\pmb{\psi}^{(0)}_{7} =
\left(
\begin{array}{c}
0 \\
1 \\
{\bf 0} \\
\imath{\boldsymbol{\mathsf M}}^{-1}\cdot{\bf N}_{\mathfrak s}
\end{array}
\right) , \qquad\ 
\pmb{\tilde\psi}^{(0)}_{7} =
\left(
\begin{array}{c}
0 \\
1 \\
{\bf 0} \\
{\bf 0}\end{array}
\right) ; 
\\
&& z^{(0)}_{8} = 0 \, , \qquad\qquad\ \ 
\pmb{\psi}^{(0)}_{8} =
\left(
\begin{array}{c}
1 \\
0 \\
{\bf 0} \\
\imath{\boldsymbol{\mathsf M}}^{-1}\cdot{\bf N}_{\mathfrak y}
\end{array}
\right) , \qquad\ 
\pmb{\tilde\psi}^{(0)}_{8} =
\left(
\begin{array}{c}
1 \\
0 \\
{\bf 0} \\
{\bf 0}\end{array}
\right) .
\eea
Accordingly, the modes $\alpha=1$-$6$ are the propagative sound modes with $z_{\alpha}^{(0)}=O(q)$, $\alpha=7$ is the heat mode, and $\alpha=8$ the vacancy diffusion mode.  Since the zeroth order is adiabatic (isoentropic), there is no damping of the modes at this approximation.

By standard perturbation theory, the first-order correction to the eigenvalues can be obtained with
\be
z^{(1)}_{\alpha} = \frac{\pmb{\tilde\psi}^{(0)\dagger}_{\alpha}\cdot{\boldsymbol{\mathsf L}}^{(1)}\cdot\pmb{\psi}^{(0)}_{\alpha}}{\pmb{\tilde\psi}^{(0)\dagger}_{\alpha}\cdot\pmb{\psi}^{(0)}_{\alpha}} \, ,
\ee
giving
\bea
z^{(1)}_{\alpha} &=& -\frac{1}{2} \Big[ - \lambda_{\alpha}^{-1/2} \left(C_{{\bf v}{\mathfrak y}}^{ca} q^c \epsilon_{\alpha}^a\right)\left(F_{{\mathfrak y}{\bf v}}^{cdb} q^c q^d \epsilon_{\alpha}^b\right) + \lambda_{\alpha}^{-1} \left(C_{{\bf v}{\mathfrak y}}^{ca} q^c \epsilon_{\alpha}^a\right)\left(D_{{\mathfrak y}{\bf u}}^{cdeb} q^c q^d q^e \epsilon_{\alpha}^b\right)  \nonumber\\
&&\quad\ - \lambda_{\alpha}^{-1/2} \left(C_{{\bf v}{\mathfrak s}}^{ca} q^c \epsilon_{\alpha}^a\right)\left(F_{{\mathfrak s}{\bf v}}^{cdb} q^c q^d \epsilon_{\alpha}^b\right) + \lambda_{\alpha}^{-1} \left(C_{{\bf v}{\mathfrak s}}^{ca} q^c \epsilon_{\alpha}^a\right)\left(D_{{\mathfrak s}{\bf u}}^{cdeb} q^c q^d q^e \epsilon_{\alpha}^b\right) \nonumber\\
&&\quad\ +  D_{\bf v}^{cadb} q^c q^d \epsilon_{\alpha}^a  \epsilon_{\alpha}^b - \lambda_{\alpha}^{-1/2} F_{{\bf v}{\bf u}}^{cadeb} q^c q^d q^e \epsilon_{\alpha}^a \epsilon_{\alpha}^b  \nonumber\\
&&\quad\ - \lambda_{\alpha}^{1/2} F_{{\bf u}{\bf v}}^{cab} q^c \epsilon_{\alpha}^a \epsilon_{\alpha}^b + D_{\bf u}^{cadb} q^c q^d \epsilon_{\alpha}^a \epsilon_{\alpha}^b\Big]  \qquad\qquad\qquad (\alpha=1,2,3) \, ;  \label{z1_123}\\
z^{(1)}_{\alpha+3} &=& -\frac{1}{2} \Big[\lambda_{\alpha}^{-1/2} \left(C_{{\bf v}{\mathfrak y}}^{ca} q^c \epsilon_{\alpha}^a\right)\left(F_{{\mathfrak y}{\bf v}}^{cdb} q^c q^d \epsilon_{\alpha}^b\right) + \lambda_{\alpha}^{-1} \left(C_{{\bf v}{\mathfrak y}}^{ca} q^c \epsilon_{\alpha}^a\right)\left(D_{{\mathfrak y}{\bf u}}^{cdeb} q^c q^d q^e \epsilon_{\alpha}^b\right)  \nonumber\\
&&\quad\ + \lambda_{\alpha}^{-1/2} \left(C_{{\bf v}{\mathfrak s}}^{ca} q^c \epsilon_{\alpha}^a\right)\left(F_{{\mathfrak s}{\bf v}}^{cdb} q^c q^d \epsilon_{\alpha}^b\right) + \lambda_{\alpha}^{-1} \left(C_{{\bf v}{\mathfrak s}}^{ca} q^c \epsilon_{\alpha}^a\right)\left(D_{{\mathfrak s}{\bf u}}^{cdeb} q^c q^d q^e \epsilon_{\alpha}^b\right) \nonumber\\
&&\quad\ +  D_{\bf v}^{cadb} q^c q^d \epsilon_{\alpha}^a  \epsilon_{\alpha}^b + \lambda_{\alpha}^{-1/2} F_{{\bf v}{\bf u}}^{cadeb} q^c q^d q^e \epsilon_{\alpha}^a \epsilon_{\alpha}^b  \nonumber\\
&&\quad\ + \lambda_{\alpha}^{1/2} F_{{\bf u}{\bf v}}^{cab} q^c \epsilon_{\alpha}^a \epsilon_{\alpha}^b + D_{\bf u}^{cadb} q^c q^d \epsilon_{\alpha}^a \epsilon_{\alpha}^b\Big]  \qquad\qquad\qquad (\alpha=1,2,3) \, ; \label{z1_456}\\
z^{(1)}_{7} &=& - D_{\mathfrak s}^{cd} q^c q^d + \left(C_{{\bf v}{\mathfrak s}}^{ca} q^c \right)\left({\boldsymbol{\mathsf M}}^{-1}\right)^{ab}\left(D_{{\mathfrak s}{\bf u}}^{cdeb} q^c q^d q^e \right)  ; \\
z^{(1)}_{8} &=& - D_{\mathfrak y}^{cd} q^c q^d + \left(C_{{\bf v}{\mathfrak y}}^{ca} q^c \right)\left({\boldsymbol{\mathsf M}}^{-1}\right)^{ab}\left(D_{{\mathfrak y}{\bf u}}^{cdeb} q^c q^d q^e \right) .
\eea
We note that, since $\lambda_{\alpha}=O(q^2)$, $\pmb{\epsilon}_{\alpha}=O(q^0)$, and ${\boldsymbol{\mathsf M}}=O(q^2)$, all the terms of these corrections are going as $z_{\alpha}^{(1)}=O(q^2)$ for $\alpha=1$-$8$.  Consequently, the corrections $z_{\alpha}^{(1)}$ and $z_{\alpha+3}^{(1)}$ with $\alpha=1,2,3$ are giving the damping rates of the propagative sound modes, while the heat mode and the vacancy diffusion modes are diffusive since $z_{7}=z_{7}^{(1)}=O(q^2)$ and $z_{8}=z_{8}^{(1)}=O(q^2)$.
Although the perturbative calculation can be continued to further corrections of $O(q^3)$, they are not relevant since the statistical-mechanical theory is limited to the leading  terms in the gradient expansion, thus, neglecting the corrections of $O(\nabla^3)=O(q^3)$.

In the case where the coefficients $F$'s are vanishing, we note that $z_{\alpha}^{(1)}=z_{\alpha+3}^{(1)}$, so that the damping rate is the same for sound modes propagating in opposite directions.  However, in crystals of the same crystallographic classes as piezoelectric crystals where the coefficients $F$'s may be non-vanishing \cite{LLv8}, these results show that the degeneracy between these damping rates may be split for sound modes propagating in opposite directions, because the terms with the coefficients $F$'s have opposite signs in $z_{\alpha}^{(1)}$ and $z_{\alpha+3}^{(1)}$.  

\section{Cubic crystals}
\label{Sec:Cubic}

Now, the dispersion relations of the eight hydrodynamic modes are analyzed in detail in the case of cubic crystals.  The consequences of the cross effects coupling the transport of momentum to those of heat and crystalline order are investigated.

\subsection{Tensors in cubic crystals}

In the case of cubic crystals, symmetric rank-two tensors are proportional to the unit tensor and symmetric rank-four tensors such as the elasticity and viscosity tensors can be expanded as
\begin{align}
	\left(\frac{\partial \sigma^{ab}}{\partial u^{cd}}\right)_{{\mathfrak s},{\mathfrak y}} &= (c_{11} + 2c_{12}) \Upsilon_0^{ab}\Upsilon_0^{cd} + (c_{11} - c_{12})\left(\Upsilon_1^{ab}\Upsilon_1^{cd}+\Upsilon_2^{ab}\Upsilon_2^{cd}\right) + 2c_{44}\left(\Upsilon_3^{ab}\Upsilon_3^{cd}+\Upsilon_4^{ab}\Upsilon_4^{cd}+\Upsilon_5^{ab}\Upsilon_5^{cd}\right) , \label{eq:Cgen} \\
	\eta^{abcd} & = (\eta_{11} + 2\eta_{12}) \Upsilon_0^{ab}\Upsilon_0^{cd} + (\eta_{11} - \eta_{12})\left(\Upsilon_1^{ab}\Upsilon_1^{cd}+\Upsilon_2^{ab}\Upsilon_2^{cd}\right) + 2\eta_{44}\left(\Upsilon_3^{ab}\Upsilon_3^{cd}+\Upsilon_4^{ab}\Upsilon_4^{cd}+\Upsilon_5^{ab}\Upsilon_5^{cd}\right) , \label{eq:etagen} 
\end{align}
with \cite{FC76}
\begin{align}
	 \Upsilon_0^{ab} & = \frac{\sqrt{3}}{3} \delta^{ab}\, , &&  \Upsilon_1^{ab} = \frac{\sqrt{6}}{2}\left(\delta^{a1}\delta^{b1}-\frac{1}{3}\delta^{ab}\right) , &&  \Upsilon_2^{ab} = \frac{\sqrt{2}}{2}\left(\delta^{a2}\delta^{b2}-\delta^{a3}\delta^{b3}\right) , \\
	 \Upsilon_3^{ab} &= \frac{\sqrt{2}}{2}\left(\delta^{a1}\delta^{b2}+\delta^{a2}\delta^{b1}\right) , && \Upsilon_4^{ab} = \frac{\sqrt{2}}{2}\left(\delta^{a1}\delta^{b3}+\delta^{a3}\delta^{b1}\right) , && \Upsilon_5^{ab} = \frac{\sqrt{2}}{2}\left(\delta^{a2}\delta^{b3}+\delta^{a3}\delta^{b2}\right) ,
\end{align}
the three isoentropic elastic constants $c_{11}$, $c_{12}$, and $c_{44}$, and the three viscosity coefficients $\eta_{11}$, $\eta_{12}$, and $\eta_{44}$, expressed in Voigt's notations.  As a consequence, we have in particular that
\be
{\boldsymbol{\mathsf M}} = (C_{{\bf v}{\bf u}}^{cadb} q^c q^d) = 
\left(
\begin{array}{ccc}
C_{11} q_x^{2} + C_{44}(q_y^{2}+q_z^{2}) & (C_{12}+C_{44}) q_{x} q_{y} & (C_{12}+C_{44}) q_{x} q_{z} \\
(C_{12}+C_{44}) q_{x} q_{y} & C_{11} q_y^{2} + C_{44}(q_x^{2}+q_z^{2}) & (C_{12}+C_{44}) q_{y} q_{z} \\
(C_{12}+C_{44}) q_{x} q_{z} & (C_{12}+C_{44}) q_{y} q_{z} & C_{11} q_z^{2} + C_{44}(q_x^{2}+q_y^{2}) \\
\end{array}
\right) 
\ee
with
\be\label{C11-C12-C44}
C_{11} \equiv \frac{c_{11}}{\rho} \, , \qquad C_{12} \equiv \frac{c_{12}}{\rho} \, , \qquad C_{44} \equiv \frac{c_{44}}{\rho} \, ,
\ee
and $q_x\equiv q^1$, $q_y\equiv q^2$, and $q_z\equiv q^3$.  Similar decompositions hold for other symmetric rank-four tensors.  In particular, the tensor~(\ref{diff-visco}) can be expressed in Voigt's notations with the three diffusivities
\be\label{Dv11-Dv12-Dv44}
D_{{\bf v}11} \equiv \frac{\eta_{11}}{\rho} \, , \qquad D_{{\bf v}12} \equiv \frac{\eta_{12}}{\rho} \, , \qquad D_{{\bf v}44} \equiv \frac{\eta_{44}}{\rho} \, ,
\ee
associated with the corresponding viscosity coefficients.

The point groups of cubic crystals are O$_{\rm h}$, O, T$_{\rm h}$, T$_{\rm d}$, and T, among which only the cubic crystals with the point groups T$_{\rm d}$ and T can accommodate piezoelectricity \cite{LLv8}.  The key is that piezoelectricity is described by a rank-three tensor, which is only non-vanishing for the classes T$_{\rm d}$ and T among cubic crystals.  In such cases, the tensors $\chi^{abc}$ and $\theta^{abc}$ are specified by a single non-vanishing coefficient because
\bea
&&\chi^{xyz}=\chi^{yzx}=\chi^{zxy}=\chi^{xzy}=\chi^{zyx}=\chi^{yxz} \equiv \chi \, , \\
&&\theta^{xyz}=\theta^{yzx}=\theta^{zxy}=\theta^{xzy}=\theta^{zyx}=\theta^{yxz} \equiv \theta \, ,
\eea
while the other coefficients are equal to zero \cite{LLv8}.  These rank-three tensors can thus be expressed as
\be
\chi^{abc}=\chi \, \Xi^{abc} \qquad \mbox{and}\qquad \theta^{abc}=\theta \, \Xi^{abc}
\ee
with
\be
\Xi^{abc} \equiv \delta^{a1}\delta^{b2}\delta^{c3}+\delta^{a2}\delta^{b3} \delta^{c1}+\delta^{a3}\delta^{b1}\delta^{c2} + \delta^{a1}\delta^{b3}\delta^{c2} +\delta^{a3}\delta^{b2} \delta^{c1}+\delta^{a2}\delta^{b1}\delta^{c3} \, .
\ee
For cubic crystals in the non-piezoelectric classes O$_{\rm h}$, O, and T$_{\rm h}$, these tensors are vanishing because $\chi=0$ and $\theta=0$.

\subsection{Hydrodynamic modes in cubic crystals}

For cubic crystals, the matrix~(\ref{L}) thus becomes
\be
{\boldsymbol{\mathsf L}}=\left(
\begin{array}{llll}
-D_{\mathfrak y} {\bf q}^2  & 
-D_{{\mathfrak y}{\mathfrak s}} {\bf q}^2 & 
- F_{{\mathfrak y}{\bf v}}^{cdb} q^c q^d & 
-\imath D_{{\mathfrak y}{\bf u}}^{cdeb} q^c q^d q^e\\
-D_{{\mathfrak s}{\mathfrak y}} {\bf q}^2  & 
-D_{\mathfrak s} {\bf q}^2  & 
-F_{{\mathfrak s}{\bf v}}^{cdb} q^c q^d  &  
-\imath D_{{\mathfrak s}{\bf u}}^{cdeb} q^c q^d q^e\\
\imath \, C_{{\bf v}{\mathfrak y}} q^a - F_{{\bf v}{\mathfrak y}}^{cad} q^c q^d &
\imath \, C_{{\bf v}{\mathfrak s}} q^a - F_{{\bf v}{\mathfrak s}}^{cad} q^c q^d & 
-D_{\bf v}^{cadb} q^c q^d & 
-C_{{\bf v}{\bf u}}^{cadb} q^c q^d -\imath F_{{\bf v}{\bf u}}^{cadeb} q^c q^d q^e\\
\imath D_{{\bf u}{\mathfrak y}} q^a &
\imath D_{{\bf u}{\mathfrak s}} q^a &
\delta^{ab} + \imath F_{{\bf u}{\bf v}}^{cab} q^c &
-D_{\bf u}^{cadb} q^c q^d 
\end{array}
\right)
\label{L-cubic}
\ee
with the following coefficients,
\bea
&& C_{{\bf v}{\mathfrak y}} \equiv -\rho^{-1} \, p_{\mathfrak y} +G_{{\bf v}{\mathfrak y}} \, , \\
&& C_{{\bf v}{\mathfrak s}} \equiv -\rho^{-1} \, p_{\mathfrak s} +G_{{\bf v}{\mathfrak s}}  \, , \\
&& D_{\mathfrak y} \equiv -\tilde{\xi} \, T_{\mathfrak y} - \tilde\zeta \, G_{{\bf v}{\mathfrak y}} = - D_{{\bf u}{\mathfrak y}} \, , \\
&& D_{{\mathfrak y}{\mathfrak s}} \equiv -\tilde{\xi} \, T_{\mathfrak s} - \tilde\zeta \, G_{{\bf v}{\mathfrak s}} = - D_{{\bf u}{\mathfrak s}}\, , \\
&& D_{{\mathfrak s}{\mathfrak y}} \equiv \tilde\kappa \, T_{\mathfrak y} + \tilde\xi \, G_{{\bf v}{\mathfrak y}} \, , \\
&& D_{\mathfrak s} \equiv \tilde\kappa \, T_{\mathfrak s} + \tilde\xi \, G_{{\bf v}{\mathfrak s}} \, ,
\eea
rank-four tensors,
\bea
&& C_{{\bf v}{\bf u}}^{abcd} \equiv -\rho^{-1} \, p_{\boldsymbol{\mathsf u}}\,  \delta^{ab}\, \delta^{cd} +G_{{\bf v}{\bf u}}^{abcd}  \, , \\
&& D_{\bf v}^{abcd} \equiv \rho^{-1} \, \eta^{abcd} \, , \\
&& D_{\bf u}^{abcd} \equiv \tilde{\xi} \, T_{\boldsymbol{\mathsf u}}\, \delta^{ab}\, \delta^{cd} + \tilde\zeta \, G_{{\bf v}{\bf u}}^{abcd} =  -D_{{\mathfrak y}{\bf u}}^{abcd} \, , \\
&& D_{{\mathfrak s}{\bf u}}^{abcd} \equiv \tilde\kappa \, T_{\boldsymbol{\mathsf u}} \, \delta^{ab}\, \delta^{cd}  + \tilde\xi \, G_{{\bf v}{\bf u}}^{abcd} \, ,
\eea
and odd-order tensors,
\bea
&& F_{{\mathfrak y}{\bf v}}^{abc} \equiv \theta\, \Xi^{abc} = - F_{{\bf u}{\bf v}}^{bac}\, , \\
&& F_{{\mathfrak s}{\bf v}}^{abc} \equiv \tilde\chi \, \Xi^{abc} \, , \\
&& F_{{\bf v}{\mathfrak y}}^{abc} \equiv \left( -\tilde\chi  \, T_{\mathfrak y} +\theta \, G_{{\bf v}{\mathfrak y}}\right) \Xi^{abc} \, , \\
&& F_{{\bf v}{\mathfrak s}}^{abc} \equiv \left( -\tilde\chi  \, T_{\mathfrak s} +\theta \, G_{{\bf v}{\mathfrak s}} \right) \Xi^{abc} \, , \\
&& F_{{\bf v}{\bf u}}^{abcef} \equiv -\tilde\chi \, T_{\boldsymbol{\mathsf u}} \, \Xi^{abc} \, \delta^{ef}  +\theta \,  \Xi^{abd} \, G_{{\bf v}{\bf u}}^{cdef} \, ,
\eea
where
\bea
&& p_{\mathfrak y} \equiv \left(\frac{\partial p}{\partial{\mathfrak y}}\right)_{{\mathfrak s},\boldsymbol{\mathsf u}} , \qquad\qquad\quad\ 
p_{\mathfrak s} \equiv \left(\frac{\partial p}{\partial{\mathfrak s}}\right)_{{\mathfrak y},\boldsymbol{\mathsf u}} , \qquad\qquad\quad\ 
p_{\boldsymbol{\mathsf u}} \equiv \frac{1}{3} \, {\rm tr} \left(\frac{\partial p}{\partial\boldsymbol{\mathsf u}}\right)_{{\mathfrak s},{\mathfrak y}} , \label{p_u}\\
&&T_{\mathfrak y} \equiv \left(\frac{\partial T}{\partial{\mathfrak y}}\right)_{{\mathfrak s},\boldsymbol{\mathsf u}} ,  \qquad\qquad\quad
T_{\mathfrak s} \equiv \left(\frac{\partial T}{\partial{\mathfrak s}}\right)_{{\mathfrak y},\boldsymbol{\mathsf u}} , \qquad\qquad\quad
T_{\boldsymbol{\mathsf u}} \equiv \frac{1}{3} \, {\rm tr} \left(\frac{\partial T}{\partial\boldsymbol{\mathsf u}}\right)_{{\mathfrak s},{\mathfrak y}} , \label{T_u}\\
&& G_{{\bf v}{\mathfrak y}}  \equiv \frac{1}{3\rho}\, {\rm tr} \left(\frac{\partial \pmb{\phi} }{\partial {\mathfrak y}}\right)_{{\mathfrak s},\boldsymbol{\mathsf u}} , \qquad G_{{\bf v}{\mathfrak s}}  \equiv \frac{1}{3\rho} \, {\rm tr}\left(\frac{\partial \pmb{\phi} }{\partial {\mathfrak s}}\right)_{{\mathfrak y},\boldsymbol{\mathsf u}} , \qquad G_{{\bf v}{\bf u}}^{abcd} \equiv \frac{1}{\rho} \left(\frac{\partial \phi^{ab}}{\partial u^{cd}}\right)_{{\mathfrak s},{\mathfrak y}} , \\
&& \tilde\xi \equiv \frac{\xi}{T} \, , \qquad\qquad\qquad \tilde\zeta \equiv \rho \, \zeta \, , \qquad\qquad\qquad \tilde\kappa \equiv \frac{\kappa}{T\rho} \, , \qquad\qquad\qquad  \tilde\chi \equiv \frac{\chi}{T\rho} \, .
\eea
We note that $p_{\boldsymbol{\mathsf u}}$ in equation~(\ref{p_u}) can be expressed in terms of the isoentropic compressibility $\kappa_{\mathfrak s}$ by $p_{\boldsymbol{\mathsf u}}=-1/\kappa_{\mathfrak s}$.  Since the compressibility is related to the elastic constants $c_{11}$ and $c_{12}$ according to $\kappa_{\mathfrak s}^{-1}=(c_{11}+2 c_{12})/3$ \cite{W98}, we have that
\be
-\rho^{-1} \, p_{\boldsymbol{\mathsf u}} = \frac{1}{\rho\kappa_{\mathfrak s}} = \frac{1}{3} \left( C_{11}+2 C_{12} \right) ,
\ee
because of equation~(\ref{C11-C12-C44}).
In this regard, the rank-four tensor $G_{{\bf v},{\bf u}}^{abcd}$ is defined with the two constants
\be
G_{11} = \frac{2}{3} \left(C_{11}-C_{12}\right) = - 2 G_{12} \qquad\mbox{and}\qquad G_{44} = C_{44} \, ,
\ee
so that we have
\bea
C_{11} &=& \frac{1}{\rho\kappa_{\mathfrak s}}  + G_{11} \, , \\
C_{12} &=& \frac{1}{\rho\kappa_{\mathfrak s}}  -\frac{1}{2}\, G_{11} \, , \\
C_{44} &=& G_{44} \, ,
\eea
as established in the thermodynamics of crystals \cite{W98}.  The equilibrium thermodynamic stability of the crystals requires the positivity of the compressibility and the specific heat:
\be
\kappa_{\mathfrak s}=-\frac{1}{p_{\boldsymbol{\mathsf u}}} > 0 \, , \qquad
{\mathfrak c}_{{\mathfrak y},\boldsymbol{\mathsf u}} \equiv T \left(\frac{\partial{\mathfrak s}}{\partial T}\right)_{{\mathfrak y},\boldsymbol{\mathsf u}} = \frac{T}{T_{\mathfrak s}}> 0 \, ,
\ee
whereupon $p_{\boldsymbol{\mathsf u}}<0$ and $T_{\mathfrak s}>0$.  Furthermore, the Gr\"uneisen parameter
\be
\gamma \equiv \frac{\rho}{T} \left(\frac{\partial T}{\partial \rho}\right)_{{\mathfrak s},{\mathfrak y}} \, ,
\ee
which is proportional to the thermal expansivity at constant pressure, is observed to be positive \cite{W98}.
Under the same circumstances, the quantity $T_{\boldsymbol{\mathsf u}}$ defined in equation~(\ref{T_u}) is thus negative,
\be
T_{\boldsymbol{\mathsf u}} = - \rho \left(\frac{\partial T}{\partial \rho}\right)_{{\mathfrak s},{\mathfrak y}} = - T \, \gamma <0\, .
\ee

Vacancies are also known as Schottky defects, in which atoms move to the crystal surface \cite{AM76,K76,IL09}.  The number of vacancies can be evaluated as $N_{\rm v}\simeq N \exp[-\beta(\varepsilon + p v_0)]$, where $N$ is the number of ions in the crystals, $\varepsilon\simeq 1$~eV the binding energy of the atoms to their lattice site, $v_0$ the volume per ion in the lattice, and $p$ the pressure.  In such a case, $N_{\rm v}\simeq 10^{-5} N$ at $T=1000$~K.  Since $\varepsilon \gg p v_0$, we have that
\be
\frac{1}{T_{\mathfrak y}} = \left(\frac{\partial{\mathfrak y}}{\partial T}\right)_{{\mathfrak s},\boldsymbol{\mathsf u}} \simeq \frac{\varepsilon}{k_{\rm B}T^2 } \, {\mathfrak y}  >0  \qquad\mbox{and}\qquad
\frac{1}{p_{\mathfrak y}} =\left(\frac{\partial {\mathfrak y}}{\partial p}\right)_{{\mathfrak s},\boldsymbol{\mathsf u}} \simeq - \frac{v_0}{k_{\rm B}T } \, {\mathfrak y} <0 \, ,
\ee
so that $T_{\mathfrak y}>0$ and $p_{\mathfrak y}<0$.  Moreover, the diffusion coefficient of vacancies is known to behave as $D_{\mathfrak y}\simeq D_{{\mathfrak y}0}\exp(-\beta E_{\rm v})$ with $D_{{\mathfrak y}0}\simeq 0.20$~cm$^2$/s and $E_{\rm v}\simeq 2.0$~eV for the diffusion of Cu in Cu \cite{K76}.  Therefore, vacancy diffusion coefficients are of the order of magnitude $D_{\mathfrak y}\sim 10^{-15}$ m$^2$/s at $T=1000$~K.  The vacancy diffusion coefficient is always positive. Since it is related by $D_{\mathfrak y} = -\tilde{\xi} \, T_{\mathfrak y} - \tilde\zeta \, G_{{\bf v}{\mathfrak y}}$ to $T_{\mathfrak y}>0$ and since the first term is expected to be larger than the second term, we should have that $\tilde\xi<0$.  The heat conductivity is also always positive $\kappa>0$.

On the one hand, experimental data are available about the elastic constants $c_{11}$, $c_{12}$, and $c_{44}$ \cite{AM76,K76,IL09}.  The sound velocities are typically of the order of a few thousand meters per second in crystals, in agreement with the values of the corresponding elastic constants and mass density. On the other hand, the experimental measurements of sound attenuation in solids are also available and they show that the diffusivities associated with the sound damping rates and due in particular to the viscosities are of the same order of magnitude as in liquids, i.e., $D_{\bf v}\sim 10^{-6}$~m$^2$/s \cite{M72}.  Heat diffusivity has a similar order of magnitude in solids \cite{L00}.

Now, our aim is to obtain illustrative examples of dispersion relations for the eight hydrodynamic modes in cubic crystals.  The $8\times 8$ matrix~(\ref{L-cubic}) is implemented in Mathematica \cite{Mathematica}.  For illustrative purposes, the value of the vacancy diffusion coefficient is taken larger than expected according to the known results reported here above.  The cases with vanishing and non-vanishing coefficients $F$'s are compared.  A first set of parameter values is chosen to plot the dispersion relations of the eight modes for vanishing coefficients~$F$'s:
\bea
&& C_{11} = 30 \, , \quad C_{12}=20 \, , \quad C_{44} = 15 \, , \nonumber\\
&& G_{{\bf v}{\mathfrak y}} = 0.5 \, , \quad G_{{\bf v}{\mathfrak s}} = - 1 \, , \nonumber\\
&& p_{\mathfrak y} =-0.2 \, , \quad p_{\mathfrak s} = 0.5 \, , \nonumber\\
&& T_{\boldsymbol{\mathsf u}} = - 2 \, , \quad T_{\mathfrak y} = 1.5 \, , \quad T_{\mathfrak s} = 1 \, , \nonumber\\
&& D_{{\bf v}11} =0.1 \, , \quad D_{{\bf v}12} =0.2 \, , \quad D_{{\bf v}44} = 0.3 \, , \nonumber\\
&& \tilde\xi = - 0.2 \, , \quad \tilde\zeta =0.1 \, , \quad \tilde\kappa = 0.1 \, , 
\label{parameters}
\eea
with
\bea
&& \tilde\chi = 0 \, , \quad \theta = 0 \, .
\label{CSGU-parameters}
\eea
The units are nm, ps, and K.
A second set of parameters is chosen for non-vanishing coefficients~$F$'s, using~(\ref{parameters}) with
\be
\tilde\chi = 0.1 \, , \quad \theta = 0.1 \, .
\label{F-CSGU-parameters}
\ee

\begin{figure}[h]
\centerline{\scalebox{0.7}{\includegraphics{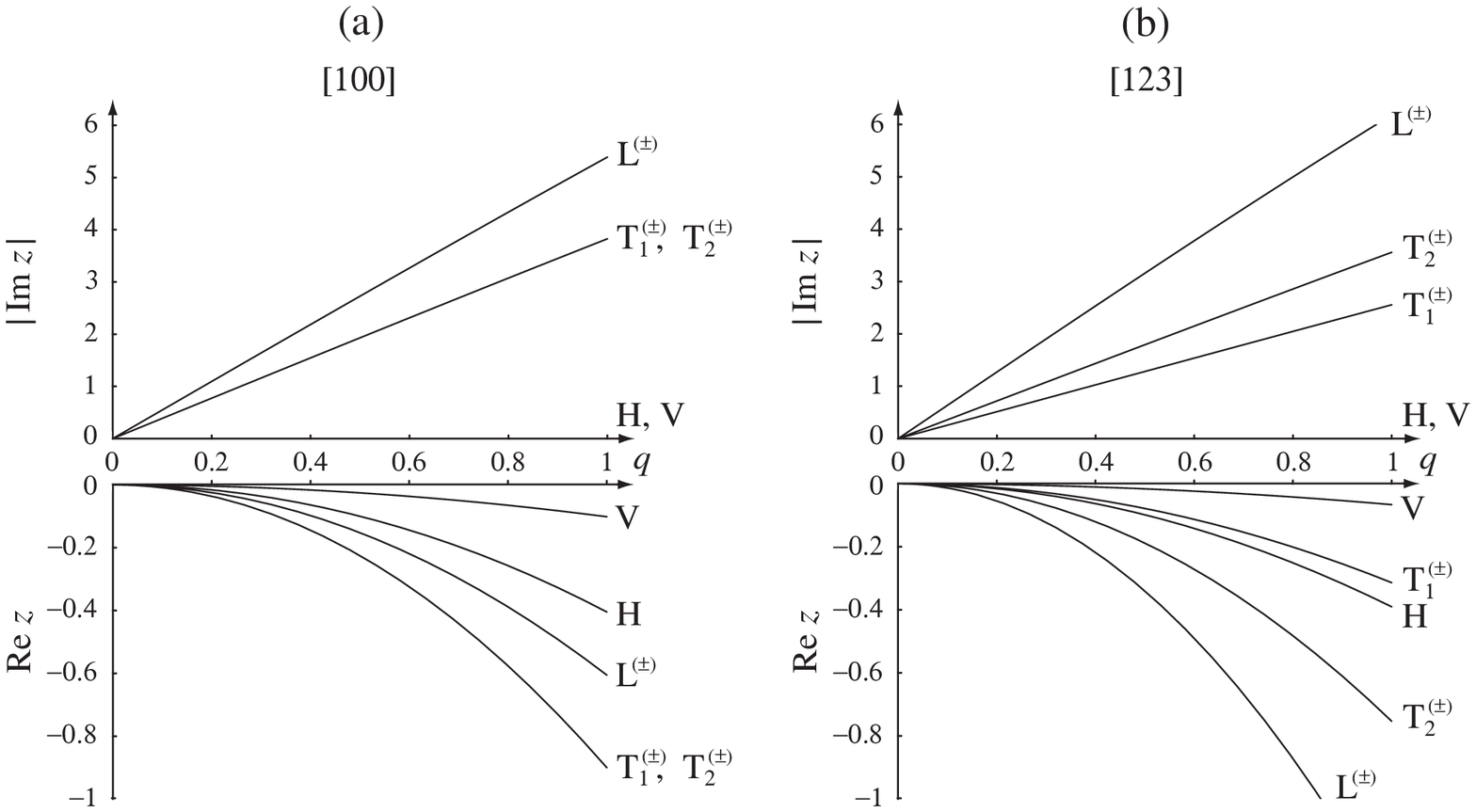}}}
\caption{Dispersion relations of the eight hydrodynamic modes in a crystal with the parameters~(\ref{parameters}) and~(\ref{CSGU-parameters}): (a) in the direction $(q_x,q_y,q_z)=(q,0,0)$; (b) in the direction $(q_x,q_y,q_z)=(q,2q,3q)/\sqrt{14}$.}
\label{fig1}
\end{figure}

\begin{figure}[h]
\centerline{\scalebox{0.7}{\includegraphics{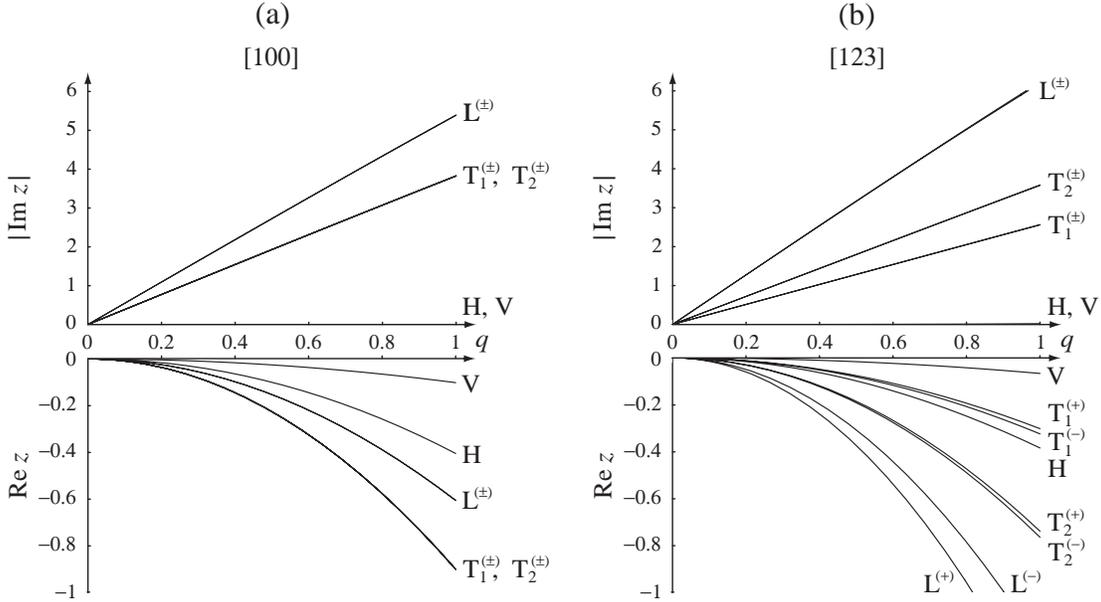}}}
\caption{Dispersion relations of the eight hydrodynamic modes in a crystal with the parameters~(\ref{parameters}) and~(\ref{F-CSGU-parameters}): (a) in the direction $(q_x,q_y,q_z)=(q,0,0)$; (b) in the direction $(q_x,q_y,q_z)=(q,2q,3q)/\sqrt{14}$.}
\label{fig2}
\end{figure}

The eight dispersion relations for the parameter values~(\ref{parameters}) and~(\ref{CSGU-parameters}) are plotted in figure~\ref{fig1}.  The modes are the two longitudinal sound modes L$^{(\pm)}$ with the largest speeds, the four transverse sound modes T$_1^{(\pm)}$ and T$_2^{(\pm)}$, the heat mode H, and the vacancy diffusion mode V.  The heat mode and the vacancy diffusion modes are always diffusive.  The mode with ${\rm Im}\, z_{\alpha}<0$ are labelled with $(+)$ and those with ${\rm Im}\, z_{\alpha}>0$ with $(-)$ in reference to the direction of propagation with respect to that of the wave vector ${\bf q}$.  The dispersion relations of the transverse sound modes T$_1^{(\pm)}$ and T$_2^{(\pm)}$ coincide in the direction $[100]$ of wave vector $(q_x,q_y,q_z)=(q,0,0)$, but they are distinct in the direction $[123]$ of wave vector $(q_x,q_y,q_z)=(q,2q,3q)/\sqrt{14}$, which is well known \cite{AM76,K76,IL09}. 

Now, for a cubic crystal in the crystallographic classes of piezoelectricity, the eight dispersion relations are plotted in figure~\ref{fig2} for the parameter values~(\ref{parameters}) and~(\ref{F-CSGU-parameters}).  The dispersion relations are the same as in figure~\ref{fig1}(a) in the direction $[100]$ of wave vector $(q_x,q_y,q_z)=(q,0,0)$.  However, they differ from those of figure~\ref{fig1}(b) in the direction $[123]$ of wave vector $(q_x,q_y,q_z)=(q,2q,3q)/\sqrt{14}$. As seen in figure~\ref{fig2}(b), the damping rates of the propagative sound modes are different for opposite propagation directions.  The results confirm the expectation from the perturbation calculation of the dispersion relations, showing that the non-vanishing coefficients $F$'s are splitting the degeneracy between the damping rates~(\ref{z1_123}) and~(\ref{z1_456}).  Accordingly, the damping of the sound modes may be different for opposite propagation directions in the crystallographic classes~(\ref{pyro})-(\ref{piezo}).


\section{Conclusion}
\label{Sec:Conclusion}

This paper applies the local equilibrium approach of reference~\cite{MG20} to crystals.  Since the three-dimensional group of space translations is broken into one of the 230 crystallographic space groups in these phases, three Nambu-Goldstone modes emerge in addition to the five slow modes associated with the conservation of mass, energy, and momentum.  As a consequence, the hydrodynamics of crystals is governed by eight slow modes.

The microscopic form of the displacement vector field, i.e., the crystalline local order parameter, is constructed on the basis of the mechanism of continuous symmetry breaking and the result previously obtained in references~\cite{SE93,S97} for cubic crystals is recovered.  The local equilibrium approach to the nonequilibrium statistical mechanics of crystals provides the microscopic expressions for the thermodynamic and transport properties of crystals.  In particular, Green-Kubo formulas are obtained for all the transport coefficients.

Because of the crystalline anisotropy, these coefficients form tensors of rank two, three, and four.  The heat conductivities form a rank-two tensor, as well as the friction coefficients of crystalline order and the coefficients of coupling between heat transport and crystalline order.  The viscosity coefficients form a rank-four tensor.  Moreover, in the same 20 crystallographic classes as those compatible with piezoelectricity, there may exist coefficients coupling the transport of momentum to those of either heat or crystalline order.  These cross effects are described by rank-three tensors.  In the 12 other crystallographic phases, these rank-three tensors are vanishing.

The hydrodynamics of crystals around equilibrium is investigated by linearizing the eight macroscopic equations ruling the local conservation laws of mass, energy, and momentum, together with the evolution equations for the three components of the displacement vector.
The eight hydrodynamic modes and their dispersion relation are obtained by solving the corresponding eigenvalue problem and by using an expansion in powers of the wave number starting from the dissipativeless crystal as a reference.  The dispersion relations are analyzed in detail for cubic crystals.

In this way, the damping rates of the hydrodynamic modes are calculated at second order in the wave number.  The cross effects existing in the 20 crystallographic classes of piezoelectricity and coupling momentum to heat or crystalline order are shown to split the degeneracy of the damping rates for the sound modes propagating in opposite generic directions.  In these crystallographic classes, the damping of sound waves may thus be stronger in some directions than in the opposite one, as a consequence of this degeneracy splitting.  Instead, in the 12 other crystallographic classes, these damping rates remain degenerate.

To conclude, we note that the fluctuating hydrodynamics of crystals can be established by adding Gaussian white noise fields $\delta{\cal J}_{c^{\alpha}}^{a}({\bf r},t)$ to the dissipative current densities ${\cal J}_{c^{\alpha}}^{a}({\bf r},t)$.  These Gaussian white noise fields should have amplitudes given by the linear response coefficients in terms of the transport coefficients according to the fluctuation-dissipation theorem \cite{LLv9,OS06}.  In this way, the effects of fluctuations on the transport properties can be studied in crystals, using the results of the present paper.

Furthermore, these results are here deduced in the framework of classical mechanics for the microscopic motion of atoms, but they can be generalized to quantum mechanics.  In particular, the classical Green-Kubo formulas can be modified into quantum-mechanical formulas by introducing the Hermitian operators for the densities and current densities \cite{M58,Z66,R66,R67,AP81} and by taking into account their non-commutativity using imaginary-time integrals up to the inverse temperature \cite{K57}.  With such considerations, the results can be extended to the dynamics of crystals in quantum regimes.


\section*{Acknowledgements}

Financial support from the Universit\'e Libre de Bruxelles (ULB) and the Fonds de la Recherche Scientifique - FNRS under the Grant PDR T.0094.16 for the project "SYMSTATPHYS" is acknowledged.


\appendix

\section{Local equilibrium mean decay rate of the order parameter}
\label{AppA}

If the crystal is close to equilibrium, the mean particle density $n_{\rm eq}({\bf r})$ has the lattice periodicity, so that it can be expanded in lattice Fourier modes as in equation~(\ref{n_eq-G}).  As a consequence, the tensor~(\ref{N-dfn}) can be expressed as
\be\label{N-G}
{\cal N}^{ab} =\sum_{\bf G} G^a G^b \vert n_{{\rm eq},{\bf G}} \vert^2 \, .
\ee
Moreover, we have that
\be\label{G-n-0}
\sum_{\bf G} G^a \vert n_{{\rm eq},{\bf G}} \vert^2 = 0 \, ,
\ee
because the following integral over a unit cell $v$ of the lattice is vanishing by the periodicity of the density
\be
\int_v n_{\rm eq} \nabla^a n_{\rm eq} \, {\rm d}{\bf r} = 0 \, .
\ee

Now, since the local equilibrium mean values of the momentum density is given by $\langle\hat{g}^c\rangle_{\rm leq}=m n_{\rm eq} v^c$, the decay rate~(\ref{micro_decay_rate}) of the crystalline order parameter has the following local equilibrium mean value
\bea
\bar{J}_{u^a}({\bf r}) &=& \langle\hat{J}_{u^a}(\mathbf{r};\Gamma)\rangle_{{\rm leq},\pmb{\lambda}}   =  - (\pmb{\cal N}^{-1})^{ab} \int {\rm d}{\bf r'} \, \Delta({\bf r}-{\bf r'})\, {\nabla'}^b n_{\rm eq}({\bf r'}) \, {\nabla'}^c\left[ n_{\rm eq}({\bf r'}) v^c({\bf r'}) \right] \nonumber\\
&=&  - (\pmb{\cal N}^{-1})^{ab} \int {\rm d}{\bf r'} \, \Delta({\bf r}-{\bf r'})\, {\nabla'}^b n_{\rm eq}({\bf r'}) \left[ {\nabla'}^c n_{\rm eq}({\bf r'})\,  v^c({\bf r'}) + n_{\rm eq}({\bf r'})\, {\nabla'}^c v^c({\bf r'})\right] \nonumber\\
&\equiv & \langle\hat{J}_{u^a}\rangle_{{\rm leq},\pmb{\lambda}}^{\rm A}  + \langle\hat{J}_{u^a}\rangle_{{\rm leq},\pmb{\lambda}}^{\rm B} \, .
\label{JA+JB}
\eea
These two terms are calculated separately using the definition~(\ref{Delta-dfn}) of the function $\Delta({\bf r}-{\bf r'})$, the expansion~(\ref{N-G}) of the periodic crystal density, and the assumption that the velocity field is a macrofield such that its Fourier modes belong to the first Brillouin zone $\cal B$.  

On the one hand, the first term in~(\ref{JA+JB}) is obtained as follows,
\bea
 \langle\hat{J}_{u^a}\rangle_{{\rm leq},\pmb{\lambda}}^{\rm A}
 &=&  - (\pmb{\cal N}^{-1})^{ab} \int {\rm d}{\bf r'} \, \Delta({\bf r}-{\bf r'})\, {\nabla'}^b n_{\rm eq}({\bf r'}) \, {\nabla'}^c n_{\rm eq}({\bf r'})\,  v^c({\bf r'}) \\
&=&  (\pmb{\cal N}^{-1})^{ab} \int {\rm d}{\bf r'} \int_{\cal B} \frac{{\rm d}{\bf k}}{(2\pi)^3} \, {\rm e}^{\imath{\bf k}\cdot({\bf r}-{\bf r'})} \sum_{{\bf G},{\bf G'}} G^b n_{{\rm eq},{\bf G}} \, {\rm e}^{\imath {\bf G}\cdot{\bf r'}} {G'}^c n_{{\rm eq},{\bf G'}} \, {\rm e}^{\imath {\bf G'}\cdot{\bf r'}} \int_{\cal B} \frac{{\rm d}{\bf k'}}{(2\pi)^3} \, {\rm e}^{\imath{\bf k'}\cdot{\bf r'}} \tilde{v}^c({\bf k'}) \nonumber\\
&=& (\pmb{\cal N}^{-1})^{ab} \sum_{{\bf G},{\bf G'}} G^b \, {G'}^c n_{{\rm eq},{\bf G}} \, n_{{\rm eq},{\bf G'}} \int_{\cal B} \frac{{\rm d}{\bf k}}{(2\pi)^3} \int_{\cal B} \frac{{\rm d}{\bf k'}}{(2\pi)^3} \, {\rm e}^{\imath{\bf k}\cdot{\bf r}}\, \tilde{v}^c({\bf k'}) \,  (2\pi)^3 \, \delta({\bf G}+{\bf G'}+{\bf k'}-{\bf k}) \, . \nonumber
\eea
Since the integrals over the wave vectors $\bf k$ and $\bf k'$ are both restricted to the first Brillouin zone $\cal B$, the Dirac delta distribution factorizes as $\delta({\bf G}+{\bf G'}+{\bf k'}-{\bf k})=\delta({\bf k}-{\bf k'})\, \delta_{{\bf G},-{\bf G'}}$.  Moreover, because of $n_{{\rm eq},-{\bf G}}=n_{{\rm eq},{\bf G}}^*$ and~(\ref{N-G}), we get
\be
 \langle\hat{J}_{u^a}\rangle_{{\rm leq},\pmb{\lambda}}^{\rm A}
 = - (\pmb{\cal N}^{-1})^{ab} \sum_{\bf G} G^b \, {G}^c \vert n_{{\rm eq},{\bf G}}\vert^2 \int_{\cal B} \frac{{\rm d}{\bf k}}{(2\pi)^3} \, {\rm e}^{\imath{\bf k}\cdot{\bf r}}\, \tilde{v}^c({\bf k}) = - (\pmb{\cal N}^{-1})^{ab} {\cal N}^{bc} \, v^c({\bf r}) = - v^a({\bf r}) \, ,
 \label{J-A}
\ee
since $(\pmb{\cal N}^{-1})^{ab} {\cal N}^{bc}=\delta^{ac}$.

On the other hand, the second term in~(\ref{JA+JB}) is vanishing, since
\bea
 \langle\hat{J}_{u^a}\rangle_{{\rm leq},\pmb{\lambda}}^{\rm B}
 &=&  - (\pmb{\cal N}^{-1})^{ab} \int {\rm d}{\bf r'} \, \Delta({\bf r}-{\bf r'})\, {\nabla'}^b n_{\rm eq}({\bf r'}) \, n_{\rm eq}({\bf r'})\, {\nabla'}^c v^c({\bf r'}) \nonumber\\
 &=&  (\pmb{\cal N}^{-1})^{ab} \int {\rm d}{\bf r'} \int_{\cal B} \frac{{\rm d}{\bf k}}{(2\pi)^3} \, {\rm e}^{\imath{\bf k}\cdot({\bf r}-{\bf r'})} \sum_{{\bf G},{\bf G'}} G^b n_{{\rm eq},{\bf G}} \, {\rm e}^{\imath {\bf G}\cdot{\bf r'}} n_{{\rm eq},{\bf G'}} \, {\rm e}^{\imath {\bf G'}\cdot{\bf r'}} \int_{\cal B} \frac{{\rm d}{\bf k'}}{(2\pi)^3} \, {\rm e}^{\imath{\bf k'}\cdot{\bf r'}} {k'}^c \, \tilde{v}^c({\bf k'}) \nonumber\\
&=& (\pmb{\cal N}^{-1})^{ab} \sum_{{\bf G},{\bf G'}} G^b \, n_{{\rm eq},{\bf G}} \, n_{{\rm eq},{\bf G'}} \int_{\cal B} \frac{{\rm d}{\bf k}}{(2\pi)^3} \int_{\cal B} \frac{{\rm d}{\bf k'}}{(2\pi)^3} \, {\rm e}^{\imath{\bf k}\cdot{\bf r}}\, {k'}^c \, \tilde{v}^c({\bf k'}) \,  (2\pi)^3 \, \delta({\bf G}+{\bf G'}+{\bf k'}-{\bf k}) \nonumber\\
&=& (\pmb{\cal N}^{-1})^{ab} \sum_{\bf G} G^b \vert n_{{\rm eq},{\bf G}}\vert^2 \int_{\cal B} \frac{{\rm d}{\bf k}}{(2\pi)^3} {\rm e}^{\imath{\bf k}\cdot{\bf r}}\, {k}^c \, \tilde{v}^c({\bf k})\nonumber\\
&=&-\imath\, (\pmb{\cal N}^{-1})^{ab} \sum_{\bf G} G^b \vert n_{{\rm eq},{\bf G}}\vert^2 \, \nabla^c  v^c({\bf r}) = 0 \, ,
\label{J-B}
\eea
because of~(\ref{G-n-0}).  Replacing the results~(\ref{J-A}) and~(\ref{J-B}) into~(\ref{JA+JB}), equation~(\ref{Ju-leq}) for the local equilibrium mean value of the decay rate is thus proved.



\end{document}